%% file: main.tex
\documentclass[11pt]{article} 
\usepackage[utf8]{inputenc}
\usepackage[T1]{fontenc}
\usepackage{tikz}
\usepackage[a4paper]{geometry}
\usepackage{microtype} 
\usepackage{lmodern}
\usepackage[labelformat=simple, labelsep=colon, justification=justified, singlelinecheck=false]{caption}
\usepackage{subcaption}
\usepackage{xcolor}
\usepackage{bigints}
\usepackage{amssymb}
\usepackage{makecell}
\usepackage{multirow} 
\usepackage{amsmath}
\usepackage{amsthm}
\usepackage{mathtools}
\usepackage{graphicx}
\usepackage{verbatim}
\usepackage{natbib}
\usepackage{bm, bbm, dsfont}
\usepackage[hidelinks]{hyperref}
\usepackage{booktabs}
\usepackage{caption}
\usepackage{rotating} 
\usepackage{float} 
\usepackage[normalem]{ulem}
\usepackage{subdepth}
\usepackage{paralist}
\usepackage{algorithm}
\usepackage{algpseudocode}
\usepackage{siunitx}
 
\definecolor{mediumpersianblue}{rgb}{0.0, 0.4, 0.65}

\newtheorem{theorem}{Theorem}[section]

\theoremstyle{definition}
\newtheorem{definition}[theorem]{Definition}
\newtheorem{example}[theorem]{Example}

\theoremstyle{remark}
\newtheorem{remark}[theorem]{Remark}

\numberwithin{equation}{section}
\input{newcommands}
\graphicspath{Fig/}

\title{A penalized least-squares estimator for extreme-value models with multiple extreme directions}

\author{
Anas Mourahib\thanks{Corresponding author. Technical University of Eindhoven, Department of Mathematics and Computer Science, Eindhoven, The Netherlands. Tel.: +31 6 347 605 62. Email: \texttt{a.mourahib@tue.nl}. Supported by the \emph{Fonds de la Recherche Scientifique – FNRS} (grant number T.0203.21), Belgium.} 
\and 
Anna Kiriliouk\thanks{UCLouvain, LIDAM/ISBA, Louvain-la-Neuve, Belgium. Email: \texttt{anna.kiriliouk@uclouvain.be}. ORCID: 0000-0003-2977-0140.} 
\and 
Johan Segers\thanks{KU Leuven, Department of Mathematics, Leuven, Belgium, and UCLouvain, LIDAM/ISBA, Louvain-la-Neuve, Belgium. Email: \texttt{jjjsegers@kuleuven.be}. ORCID: 0000-0002-0444-689X.}
}

\date{\today}

\begin{document}

\maketitle

\begin{abstract}
Estimating the parameters of max-stable parametric models poses significant challenges, particularly when some parameters lie on the boundary of the parameter space. This situation arises when a subset of variables exhibits extreme values simultaneously, while the remaining variables do not---\amtwo{a phenomenon commonly referred to as an extreme direction. A novel estimator is proposed for the parameters of a general parametric mixture model, incorporating a threshold exceedances approach based on a pseudo-norm penalization. The latter plays a crucial role in accurately identifying parameters at the boundary of the parameter space. Additionally, the estimator comes with a data-driven algorithm to detect groups of variables corresponding to extreme directions. The performance of the estimator is assessed in terms of both parameter estimation and the identification of extreme directions through extensive simulation studies. Finally, the method is applied to two real-world datasets: discharge measurements at stations along the Danube river, and financial portfolio losses from stocks listed on the NYSE, AMEX, and NASDAQ. In both applications, the sets of variables that can become large simultaneously are identified.}
\end{abstract}

\noindent%
\amtwo{{\it Keywords:}  extreme directions, max-linear model, penalized least-squares, stable tail dependence function, threshold exceedances}

\section{Introduction}
\label{sec:introduction}
Consider a random vector $\bX=(X_1,\ldots , X_d)$ of risk factors and suppose the focus is on rare events.  In this regime, neither parametric nor nonparametric methods are satisfactory: parametric models rely on distributional assumptions that can be dominated by the bulk of the data, and thus tend to underestimate extreme outcomes, while nonparametric techniques demand large sample size, which are often unavailable when studying extremes.  Extreme value theory offers a rigorous framework for extrapolating beyond observed levels and has found applications in finance, hydrology, weather forecasting, and other fields.

Univariate extreme value theory, that is, when $d=1$, has been extensively studied \citep{haan2006extreme,Be04,coles2001introduction} and modeling rare events in this case is done via a finite‐dimensional parametric family of distributions.
Multivariate modeling of $d \geq 2$ risk factors is a more challenging topic, since  risks may exhibit some sort of dependence.  The goal is not only to model rare events of each risk factor independently but to capture the extremal dependence between these risk factors.  Two main approaches appear in this context: the block‐maxima approach and the threshold exceedances approach. In the first approach, standardized componentwise maxima of $\bX$ are asymptotically modeled using a max‐stable random vector $\bZ=(Z_1,\ldots , Z_d)$.  In the second approach, asymptotically, observations where at least one risk factor is extreme are modeled using a multivariate (generalized) Pareto random vector $\bY=(Y_1,\ldots, Y_d)$ \citep{rootzen2006multivariate}.

For both approaches, many parametric families have been proposed; popular examples include the Hüsler–Reiss model \citep{hueslerReiss1989} and the logistic model \citep{T1990}, the latter being a special case of the scaled extremal Dirichlet model \citep{belzile2017extremal}. Estimation methods for such models are numerous, for example, through censored likelihood \citep{padoan2010likelihood}, weighted least squares \citep{einmahl2018continuous}, or based on neural networks \citep{lenzi2023neural,richards2024neural}.

\begin{figure}
    \centering
    \begin{minipage}[t]{0.45\textwidth}
        \includegraphics[width=0.9\textwidth]{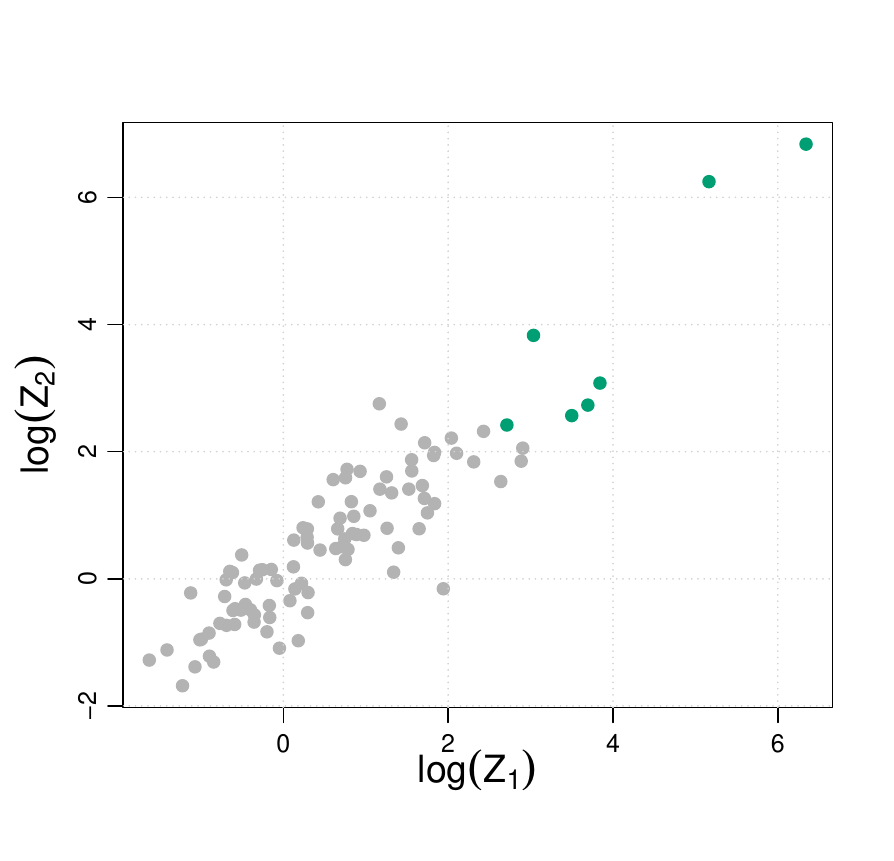} 
    \end{minipage}%
    \begin{minipage}[t]{0.45\textwidth}
        \includegraphics[width=0.9\textwidth]{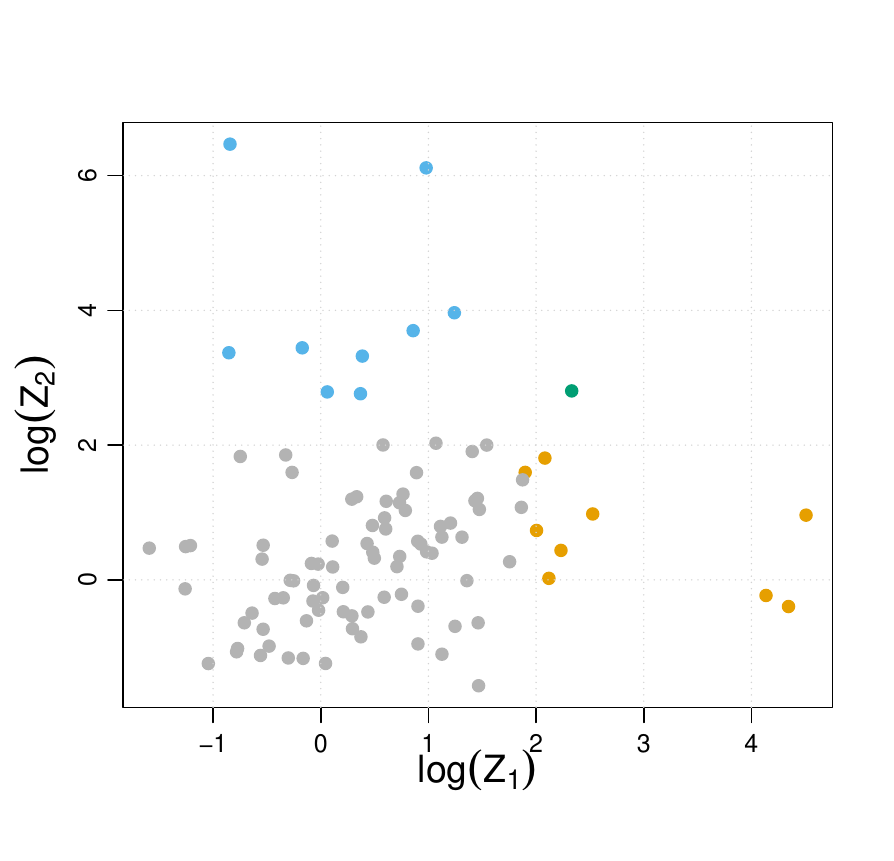} 
    \end{minipage}

    \caption{\label{fig:HRandlogistic}\amtwo{A sample of size $n = 100$ from a max-stable \HR~model (left) and a max-stable mixture \HR~model (right). Points in the extreme direction $\{1,2\}$ are shown in green, those in $\{1\}$ in orange, and those in $\{2\}$ in blue.}
    }

\end{figure}

\amtwo{A limitation of many parametric models is that they enforce all components} of $\bX$ to be extreme simultaneously. Figure~\ref{fig:HRandlogistic} (left) shows a log-transformed simulation from a max-stable \HR~model.
Observations are close to the diagonal line \(x=y\), suggesting that both components are large simultaneously. However, in practice, it may happen that only certain subsets $J\subseteq\{1,\dots,d\}$ of variables are jointly extreme while the others are not. Such a subset $J$ is called an \emph{extreme direction} \citep{mourahib2024multivariate, simpson2020determining}.

\amtwo{Multiple extreme directions can be accommodated in the max-linear model~\citep{F12,Ei12} and the asymmetric logistic model~\citep{T1990}}. However, the first model has a singular distribution (\amtwo{neither} discrete nor absolutely continuous), making statistical inference complicated, while the second one \amtwo{may be too specific to represent a broad class of max-stable dependence structures. Moreover, existing estimation methods for these models are not designed to identify extreme directions, as this requires reliably detecting parameters on the boundary of the parameter space.} 

To capture multiple extreme directions, \citet{mourahib2024multivariate} introduce the \emph{mixture model} that extends the classic max‐linear model and can represent \emph{any} max‐stable distribution.  It is identified via:
\begin{compactitem}
  \item a  $(d \times r)$ coefficient matrix $A$  with entries in $[0,1]$, whose columns correspond to distinct extreme directions of $\bX$.  Zero entries in $A$ indicate which variables are not contained in a given extreme direction \citep[Proposition~4.5]{mourahib2024multivariate};
  \item a second set of parameters that governs the dependence structure \emph{within} each extreme direction.
\end{compactitem}
\amtwo{A simulated sample from the mixture model is shown in Figure~\ref{fig:HRandlogistic} (right), where we can see that it is capable of not only representing the extreme direction $\{1,2\}$ (points in black), but also $\{1\}$ and $\{2\}$ (points in orange and blue, respectively).} 

A natural question is how to estimate the model parameters. The weighted least‐squares estimator of \citet{einmahl2018continuous} requires the true parameters to be in the interior of the parameter space.  In contrast, the mixture model’s reliance on zero entries in $A$ pushes parameters to the boundary of the space, violating this condition. We propose an estimator based on threshold exceedances for the parameters of max-stable distributions that accommodate arbitrary collections of extreme directions, building on the least squares estimator introduced in~\cite{einmahl2018continuous}. A pseudo-norm $L_p$ with $p\leq 1$ (see Section~\ref{sec:estimation_mixture_model}) is used to penalize the coefficient matrix $A$ of the mixture model, which leads to zero entries in $A$ and thus multiple extreme directions, in particular those different from $\cbr{1,\ldots,d}$. The proposed penalized least squares estimator can simultaneously identify extreme directions \textit{and} estimate the mixture model parameters. This sets our approach apart from the existing literature, which focuses exclusively on identifying extreme directions~\citep{goix2017JMVA,simpson2020determining,meyer2020multivariate}. 

A common criticism of extreme-value models possessing asymptotic dependence is their difficulty in handling situations where not all variables are extreme, that is, when the extreme directions differ from the full set $\{1, \ldots, d\}$. This issue has motivated alternative approaches such as conditional extremes~\citep{heffernan2004conditional} and geometric extremes~\citep{wadsworth2024statistical}. \amtwo{We address this limitation via our algorithm 
that can detect multiple extreme directions}, including those involving only subsets of variables. We compare our approach with the DAMEX algorithm from~\citep{goix2017JMVA}, which considers a subset $J \subseteq \{1,\ldots,d\}$ to define an extreme direction if the corresponding cone $\mathcal{E}_J = \{ \bm{x} \geq 0 : x_j > 0 \text{ iff } j \in J \}$ has positive mass under an empirical estimate of the exponent measure~\citep[Equation~20]{goix2017JMVA}. However, as noted in~\citet{wadsworth2024statistical}, at finite levels, data rarely lie exactly on such cones when $J \neq \{1,\ldots,d\}$, which can lead to missed directions. In contrast, our algorithm uses tuning parameters that can be selected via cross-validation, improving robustness and practical applicability.



The paper is structured as follows. Section~\ref{sec:background} provides essential background on extreme value theory, focusing in particular on the notions of extreme directions~\citep{simpson2020determining}, the mixture model~\citep{mourahib2024multivariate} and the least-squares estimator~\citep{einmahl2018continuous}. In Section~\ref{sec:estimation_mixture_model}, we introduce a penalized estimator for the mixture model, \amtwo{prove its consistency}, and develop a data-driven algorithm to identify the extreme directions present in a given sample. The simulation study in Section~\ref{sec:simu_study} illustrates that the choice of the tuning parameters in our estimation procedure effectively reduces to the choice of a single penalization parameter, which we determine via cross-validation. Moreover, with this choice of tuning parameter, we show that our estimation procedure gives good results both in terms of estimation of the mixture model and extreme directions identification; \amtwo{in particular, it outperforms the DAMEX algorithm}. Finally, in Section~\ref{sec:application}, we illustrate the practical relevance of our approach by applying the algorithm, first to river discharge data across stations from the Danube basin, and second to financial industry portfolios constructed from stocks listed on the NYSE, AMEX, and NASDAQ. 

\paragraph{Notation.} Let $d$ be a positive integer and write $D = \cbr{1,\ldots,d}$. Throughout, bold symbols will refer to multivariate quantities. Let $\bzero$ and $\bone$ denote vectors of zeroes and ones, respectively, of a dimension clear from the context. Let $\bx=(x_1,\ldots,x_d)$ denote a $d$-dimensional vector and, for a non-empty set $J \subseteq \cbr{1,\ldots,d}$, write $\bx_J=(x_j)_{j \in J}$. Mathematical operations on vectors such as addition, multiplication and comparison are considered component-wise. For two vectors $\bo{a}, \; \bo{b}$, the set $[\bo{a},\bo{b}]$ will denote the Cartesian product $\prod_{j=1}^{d}[a_j,b_j]$ and so forth for $(\bo{a},\bo{b}), \; [\bo{a},\bo{b}), \; (\bo{a},\bo{b}]$.
\amtwo{For two sets $A , B $, we define their symmetric difference by $A \symdif B = (A \setminus B ) \cup (B \setminus A)$. For two non-negative integers $d, r$, let $[0,1]^{d \times r}$ denote the set of $d \times r$ matrices with entries in $[0,1]$.}
The arrow $\dto$ denotes convergence in distribution (weak convergence). For two vectors $\bm{x}$ and $\bm{y}$ in $\mathbb{R}^d$, define the \emph{lexicographic order} $\leq_{\text{lex}}$ as 
\begin{equation}
\label{lexi_order}
\bm{x} \leq_{\text{lex}} \bm{y} \quad \text{if and only if} \quad x_j < y_j \text{ for the first index } j \in D \text{ for which } x_j \neq y_j.
\end{equation}
The lexicographic order is a total order in $\mathbb{R}^d$, meaning that either $\bx \leq_{\text{lex}} \by$ or $\by \leq_{\text{lex}} \bx$. For a set $S$, let 
$S^n = S \times \dots \times S$ ($n$ times). Finally, let $\EE := [0,\infty)^d\setminus \cbr{\bzero}$. 

\paragraph{Implementation.} The source code used to conduct the experiments and generate the results in this paper is available at \url{https://github.com/AnasMourahib/Penalized_least-squares_estimator}, providing full transparency and supporting reproducibility.

\section{Background}
\label{sec:background}

\subsection{Multivariate extreme value distributions} \label{09:12:15:48}

For $i = 1, \ldots, n$, let $\bX_i = \rbr{X_{i1},\ldots, X_{id}}$ be iid\ copies of $\bX$, a random vector on $\Rd$ with joint cumulative distribution function (cdf) $F$ and margins $F_1,\ldots,F_d$. We say that $F$ is in the (max-)domain of attraction of a \emph{max-stable} distribution $G$ 
if there exist sequences $\boldsymbol{a}_{n} \in (0,\infty)^d$ and $\boldsymbol{b}_{n} \in \Rd$ such that
\begin{equation*}
    \dfrac{\max_{i=1,\ldots,n}\bX_{i}- \boldsymbol{b}_{n}}{\boldsymbol{a}_{n}} \dto G, \qquad n \to \infty.
\end{equation*}
The margins of $G$, denoted $G_1,\ldots,G_d$, are univariate extreme-value distributions themselves, 
and any max-stable distribution $G$ admits the representation
  \begin{equation*}
    G(\bx)= \exp \cbr{  - \ell \rbr{ - \ln G_1(x_1), \ldots, - \ln G_d(x_d)}},  
  \end{equation*}  
where $\ell:[0,\infty)^d \rightarrow [0, \infty)$ is the \emph{stable tail dependence function (stdf)}. Then the distribution of $\bX^*
 = (X_1^*,\ldots,X_d^*)$ with $X_j^* = 1/\cbr{1-F_{j}(X_{j})}$ for $j \in D$ is in the domain of attraction of $G^*(\bx) = \exp \cbr{- \ell \rbr{1/x_1,\ldots,1/x_d}}$, a max-stable distribution with unit-\FR{} margins.


The stable tail dependence function can be retrieved via
\begin{equation}\label{eq:stdf}
    \ell(\bx) = \lim_{t \downarrow 0} t^{-1} \PP \sbr{ F_1(X_{1}) \geq 1 - tx_1 \text{ or } \ldots \text{ or } F_d(X_{d}) \geq 1 - tx_d}.
\end{equation}

The function $\ell$ is convex and homogeneous in the sense that for $c \ge 0$ and $\bx \in [0,\infty)^d$, we have $\ell(c\bx)=c\ell(\bx)$. Moreover, unit-\FR{} margins of $G^*$ imply that $\ell(0,\ldots,0,x_j,0,\ldots,0)=x_j$ for $x_j \ge 0$ and $j \in D$. Finally, $\ell$ can be bounded by
\begin{equation}\label{eq:ellbound}
\max(x_1, \ldots, x_d) \leq \ell(x) \leq x_1 + \cdots + x_d,
\end{equation}
where the lower bound corresponds to perfect tail dependence and the upper bound to asymptotic independence.

Suppose that $\bZ = (Z_1,\ldots,Z_d)$ follows a max-stable distribution with unit-Fr\'echet margins.  For non-empty $J \subseteq D$, let $\ell_{J}$ be the stdf associated with $\bZ_{J}$, that is,
\begin{equation*}
  \ell_{J} (\bx_{J})= \ell \rbr{\Tilde{\bx}}, 
  \qquad \bx_{J} \in [0, \infty)^{\lvert J \rvert},
\end{equation*}
with $\Tilde{\bx}= \sum_{j \in J} x_j \bo{e}_j$ and $\bo{e}_j$ the $j$-th canonical unit vector in $\Rd$. 
The \emph{extremal coefficients} \citep{schlather2002inequalities} are defined via $\zeta_J := \ell (\sum_{j \in J} \bo{e}_j)$, for sets $J \subseteq D$ with $|J| \geq 2$. 
From~\eqref{eq:ellbound}, we get that $1 \leq \zeta_J \leq |J|$. The value $\zeta_J$ can be interpreted as the effective number of tail independent variables among $\bZ_J$.

The \emph{exponent measure} $\mu$ of $G^{*}$ is the unique Borel measure concentrated on $\EE = [0,\infty)^d\setminus \cbr{\bzero}$ such that 
\begin{equation*}
 \mu\rbr{\EE \setminus \left[\bzero,1/\bx\right]} = \ell (\bx), 
    \qquad  \bx \in [\bzero, \binfty)^d.
\end{equation*}
The measure $\mu$ is finite on Borel sets that are bounded away from the origin. Moreover, $\mu\rbr{\EE \setminus \left[\bzero,\by\right]} = \infty$ as soon as $y_j=0$ for some $j \in D$.


\subsection{Extreme directions and the mixture model}\label{subsec:ed}
For non-empty $J \subseteq D$, define $\EE_J:=\cbr{\bx \in \EE: \ x_j>0 \text{ iff } j \in J}$. Then $\cbr{ \EE_J : \; \varnothing \ne J \subseteq D}$ forms a partition of $\EE$. Let $\bX^*$ be again a $d$-dimensional random vector in the domain of attraction of a max-stable distribution with unit-Fr\'echet margins and exponent measure $\mu$. 
The following definition was introduced in \citet{simpson2020determining} in order to identify \emph{extreme directions} of $\bX^*$. Intuitively, an extreme direction is a group of components of $\bX^*$ that may take on large values simultaneously while the other components are of smaller order.

\begin{definition}
The set $J$ is said to be an extreme direction of $\bX^*$ or $\mu$ if $\mu(\EEJ)>0$.
\end{definition}

Note that $\mu$ has no mass outside $\bigcup_{J \in \mathcal{R}(\mu)} \EE_J$, where $\mathcal{R}(\mu)$ is the collection of extreme directions of $\mu$. 
Further, even if a certain set $J \subseteq D$ is not an extreme direction of $\bX^*$, it is still possible that $J$ is a superset or a subset of an extreme direction. \amtwo{Finally, note that every margin $j$ appears in at least one extreme direction $J \in \mathcal{R}(\mu)$.}


Many standard max-stable models are only appropriate for data exhibiting a single extreme direction $J = D$. Notable exceptions are the asymmetric logistic \citep{tawn1988bivariate} and the max-linear \citep{wang2011conditional,Ei12} models.   
\cite{mourahib2024multivariate} introduce the \emph{mixture model}, a smoothed version of the max-linear model, which allows any collection  
of non-empty subsets of $D$ that together covers $D$, so that no element of $D$ is left out, to be the extreme directions of $\mu$. The mixture model is key to modeling data with non-standard dependence structures as it covers \emph{all} max-stable distributions.

\begin{definition}[\citet{mourahib2024multivariate}]
\label{def:mixture}

Let $r$ be a positive integer and write $R=\{1,\ldots,r\}$. Let $\bo{Z}_{\bigcdot 1},\ldots,\bo{Z}_{\bigcdot r}$ be independent random vectors in $\Rd$. 
For all $s \in R$, suppose that $\bo{Z}_{\bigcdot s} = (Z_{1s}, \ldots, Z_{ds})$ are max-stable random vectors with unit-Fr\'echet margins, stdf $\ell^{(s)}$ and a single extreme direction $D$. \amtwo{Let $A = (a_{js})_{j \in D;s \in R}$ be an element of $\Theta_A$, the set of $d \times r$ matrices with elements $a_{js}$ in $[0, 1]$, unit row sums $\sum_{s=1}^{r} a_{js}=1$ for all $j \in D$ and positive column sums $\sum_{j=1}^d a_{js} > 0$ for all $s \in R$.} 
The $d$-variate random vector
\begin{equation}
  \label{mixturemodel}
    \bM 
    = (M_1, \ldots, M_d) 
    = \left(\max_{s \in \bigar}\{a_{1s} Z_{1s}\}, \ldots, \max_{s \in \bigar}\{a_{ds} Z_{ds}\}\right),   
\end{equation}
will be called the $d\times r$ \emph{mixture model} with coefficient matrix $A$ and factors $\bo{Z}_{\bigcdot 1} ,\ldots,\bo{Z}_{\bigcdot r}$.
\end{definition}

\begin{remark}
Note that two different coefficient matrices, containing the same columns but in different orders, will lead to the same mixture model $\bM$ if the distributions of $\bo{Z}_{\bigcdot 1}, \ldots, \bo{Z}_{\bigcdot r}$ are identical. To ensure identifiability, we will always assume that the columns of $A$ are decreasingly ordered according to the lexicographic order~\eqref{lexi_order}.
\end{remark}

In \eqref{mixturemodel}, for a given $j \in D$, only those $s \in R$ that satisfy $a_{js} > 0$ matter. Let 
\begin{equation}
\label{eq:Jk}
    J_s := \cbr{ j \in D: \; a_{js} > 0 }, 
\end{equation}
denote the $s$-th \emph{signature} of the coefficient matrix $A$.

\begin{theorem}[\citet{S03,mourahib2024multivariate}]
\label{stephenson}
  The random vector $\bM$ in \eqref{mixturemodel} follows a max-stable distribution with unit-Fr\'echet margins and stdf 
\begin{equation}
\label{stdfgeneralmodel}
    \ell(\bx) = \sum_{s \in R} \ell^{(s)}_{J_s}\rbr{\rbr{a_{js}x_j}_{j \in J_s}},
    \qquad \bx \in [0, \infty)^d.
\end{equation}
\end{theorem}

Zero entries of the coefficient matrix $A$ allow some groups of components of the random vector $\bM$ to constitute an \extreme. \citet[Proposition~4.5]{mourahib2024multivariate} show that the set of extreme directions of the mixture model is given by the signatures of its coefficient matrix $A$. The model is general in the sense that any max-stable distribution with unit-\FR~margins is the cdf of a mixture model whose coefficient matrix has mutually different signatures \citep[Theorem~4.8]{mourahib2024multivariate}.

\begin{example}[Mixture logistic model]
\label{example:mix_log}
    For $s \in R$, suppose that the stdf associated with $\bZ_{J_s, s}$ is
\begin{equation*}
    \ell_{J_s}^{(s)}(x_1,\ldots,x_{\lvert J_s \rvert})
    = \rbr{x_1^{1/\alpha_s} + \cdots +x_{\lvert J_s \rvert}^{1/\alpha_s}}^{\alpha_s},
\end{equation*}
for $\bx \in [0, \infty)^{|J_s|}$, where $0 < \alpha_s < 1$. The stdf of the \emph{mixture logistic model} $\bM$ is
\begin{equation*}
    \ell(\bx) = \sum_{s \in R} \cbr{ \sum_{j  \in J_s} \rbr{a_{js}x_j}^{1/\alpha_s} }^{\alpha_s},
    \qquad \bx \in [0, \infty)^d.
\end{equation*}
\amtwo{In the rest of the paper, we will  assume that each extreme direction only corresponds to a single column of the coefficient matrix $A$, i.e., we will assume that the signatures of $A$ are mutually different. Under this assumption, the mixture logistic model coincides with the asymmetric logistic model.} 
\end{example}

\begin{example}[Mixture \HR~model]
\label{example:mix_HR}
    Let $s \in R$. 
    For $\lvert J_s \rvert >1$, suppose that $\ell_{J_s}^{(s)}$ is the \HR~stdf associated with the $\rbr{\lvert J_s \rvert \times  \lvert J_s \rvert}$ variogram matrix\footnote{A matrix $\Gamma \in [0,\infty)^{d \times d}$ is a variogram matrix if it is symmetric, has zero diagonal, and satisfies $\bo{v}^{\top} \Gamma \bo{v}<0$ for all $\bzero  \neq \bo{v} \perp \bone$. To each positive definite covariance matrix $\Sigma \in \reals^{d \times d}$ is associated a variogram matrix $\Gamma$ via $\Gamma_{st}=\Sigma_{ss}+\Sigma_{tt}-2 \Sigma_{st}$ for $s,t \in D$.} 
    $\Gamma^{(s)}$; see, e.g., \citet{huser2013composite}. That is, 
\[ \ell_{J_s}^{(s)}(\bx)=
\sum_{ j \in J_s} x_j \, \Phi_{\lvert J_s \rvert-1}\rbr{\bo{\eta}^{j,(s)}\rbr{\bx}; \Sigma^{j,(s)} }, \qquad \bx \in [0,\infty)^{J_s},
\]
where for $j \in J_s$, 
$\bo{\eta}^{j,(s)}\rbr{\bx} := \cbr{\ln(x_j/x_t)+\Gamma^{(s)}_{jj}/2}_{t \in J_s; t \neq j}$ 
where by convention, $0/0$ is set to $\infty$, and 
where the $\rbr{\lvert J_s \setminus \{j\} \rvert \times  \lvert J_s\setminus \{j\} \rvert}$ matrix $\Sigma^{j,(s)}$ is defined by
\begin{equation*}
        \Sigma^{j,(s)}: =\frac{1}{2} \cbr{\Gamma_{jt}^{(s)}+\Gamma_{jt'}^{(s)}-\Gamma_{tt'}^{(s)}}_{t,t' \in J_s \setminus \{j\}},
        \label{equa:matrix} 
\end{equation*}
and finally, where $\Phi_{m}( \point;\Sigma)$ is the cdf of the centered $m$-variate normal distribution  with covariance matrix $\Sigma$. The stdf of the mixture model $\bM$ is 
\begin{equation*}
    \ell(\bx) = \sum_{s \in R} \cbr{ \sum_{ j \in J_s } a_{js} x_j \, \Phi_{\lvert J_s \rvert-1}\rbr{\bo{\eta}^{j,(s)}\rbr{\rbr{a_{js}x_j}_{j \in J_s}}; \Sigma^{j,(s)} } },
    \qquad \bx \in [0, \infty)^d.
\end{equation*}
\end{example}

\subsection{Least squares estimator of tail dependence}
Let again $\bX_1, \dots, \bX_n$ denote iid random vectors in the domain of attraction of a max-stable distribution with stdf $\ell$. We will first recall how to estimate $\ell$ non-parametrically. Let $k = k_n \in (0,n]$ 
be an intermediate sequence such that 
\begin{equation}
\label{equa:exc_prop}
        k \to \infty \quad \text{and} \quad k/n \to 0, \quad \text{as } n \to \infty.
\end{equation}
\amtwo{The \emph{tail fraction} $k/n$ determines the fraction of the data that is considered as extreme and that will be used for inference.} A standard non-parametric estimator of the stdf is 
 \begin{equation}
\label{eq:ellclassic}
  \hat{\ell}_{n,k} (\bx) = 
  \frac{1}{k} \sum_{i=1}^n
  \mathbbm{1}
  \left\{
    R_{i1}^n > n + 1/2 - kx_1  \text{ or } \ldots \text{ or }  R_{id}^n > n + 1/2 - kx_d
  \right\},
\end{equation}
where $R_{ij}^n$ denotes the rank of $X_{ij}$ among $X_{1j}, \ldots , X_{nj}$ for $j=1,\ldots,d$. Note that \eqref{eq:ellclassic} is a minor modification of the standard empirical stdf \citep{drees1998best}. Alternatives are for example the bias-corrected estimator proposed in \citet{beirlant2016bias} or a smooth estimator based on the empirical beta copula \citep{kiriliouk2018estimator}.

Suppose now that $\ell$ belongs to a parametric family $\cbr{\ell(\, \cdot \ ; \bo{\theta}): \bo{\theta} \in \Theta }$, with $\Theta \subseteq \reals^{p_{\Theta}}$ for some integer $p_{\Theta} \geq 1$. Furthermore, assume the existence of a unique parameter $\bo{\theta}_0 \in \Theta$ such that $\ell(\bx) = \ell(\bx; \bo{\theta}_0)$ for all $\bx \in [0,\infty)^d$.
\citet{einmahl2018continuous} propose a continuous updating weighted least-squares estimator for $\bo{\theta}_0$, which, in its simplest form, reduces to the ordinary least-squares estimator
\begin{equation}
    \label{equa:unpen_lss}
    \hat{\bo{\theta}}_{n,k} := \argmin_{\bo{\theta} \in \Theta} \sum_{m=1}^q \rbr{\hat{\ell}_{n,k}(\bo{c}_m) - \ell(\bo{c}_m; \bo{\theta}) }^2. 
\end{equation}
Here, $\bc_1,\ldots,\bc_q \in [0,\infty)^d$, where $\bc_m = (c_{m1},\ldots,c_{md})$ for $m = 1,\ldots,q$ are the $q$ points in which $\ell$ and $\hat{\ell}_{n,k}$ are evaluated. These grid points
\(\bm{c}_1,\ldots,\bm{c}_q\) must be chosen such that the mapping $L : \Theta \to \mathbb{R}^q, \,
\bo{\theta} \mapsto \{\ell(\bm{c}_m;\bo{\theta}),\; m=1,\ldots,q\},$
is one-to-one, ensuring that the parameter vector \(\bo{\theta}\) is identifiable from
\(\{\ell(\bm{c}_m;\bo{\theta}), m=1,\ldots,q\}\); see
\citet[Section~2.2]{einmahl2018continuous}. The estimator \eqref{equa:unpen_lss} is easy to compute and shows good performance for many well-known max-stable models \citep[Section 3]{einmahl2018continuous}. 
\amtwo{However, it is not suitable for the mixture model, where some parameters may lie on the boundary of the parameter space. An extension is proposed in~\citet{kiriliouk2020hypothesis}, but unfortunately, it is based on conditions that may be difficult to verify}.

\section{Estimation of the mixture model}
\label{sec:estimation_mixture_model}

In Section~\ref{sec:construction}, we will start by introducing our estimation procedure for the simplified setting where the number of extreme directions, $r$, is assumed to be known. The case where $r$ is unknown will be discussed in Section~\ref{sec:EDI_algo}.

\subsection{Known number of extreme directions}
\label{sec:construction}

As in Definition~\ref{def:mixture}, let $\bM$ be the $d \times r$ mixture model with coefficient matrix $A$ and factors $\bo{Z}_{\bigcdot 1},\ldots,\bo{Z}_{\bigcdot r}$ with stdfs $\ell^{(1)},\ldots, \ell^{(r)}$. Recall $R= \cbr{1,\ldots, r}$.
 From now on, suppose that the stdfs \( \ell^{(1)}, \ldots, \ell^{(r)} \) belong to a common parametric family \( \{ \ell(\,\cdot \ ; \bo{\theta}_Z) : \bo{\theta}_Z \in \Theta_Z \} \), with \( \Theta_Z \subseteq \mathbb{R}^v \) for some integer $v \geq 1$, and that for each $s \in R$, there exists a unique parameter vector $\bo{\theta}^{(s)}_Z \in \Theta_Z$  such that  
$
\ell^{(s)} = \ell( \, \cdot \ ; \bo{\theta}_Z^{(s)})
$. In what follows, for simplicity, we assume that the dependence parameters across the columns are identical, meaning that all $\boldsymbol{\theta}_Z^{(s)}$, $s \in R$, are equal to a common $\boldsymbol{\theta}_Z \in \Theta_Z$. However, all the results remain valid in the general case.

\aktwo{For $\bo{\theta} = (A, \bo{\theta}_Z) \in [0, 1]^{d \times r} \times \Theta_Z$, define $\ell(\cdot\,;\bo{\theta}) : [0, \infty)^d \to [0, \infty)$ by
\[
    \ell(\bx, \btheta)
    = \sum_{s \in R} \ell_{J_s}\rbr{\rbr{a_{js}x_j}_{j \in J_s}; \btheta_Z},
    \qquad \bx \in [0, \infty)^d.
\]
Recall $\Theta_A \subset [0, 1]^{d \times r}$ in Definition~\ref{def:mixture}. If $A \in \Theta_A$, then $\ell(\cdot\,, \btheta)$ is the stdf of the mixture model $\bM$ in Theorem~\ref{stephenson}, but $\ell(\cdot\,, \btheta)$ is defined for $A \in [0, 1]^{d \times r} \setminus \Theta_A$ too.} 
\aktwo{We use the extended parameter space $[0, 1]^{d \times r}$ in the preliminary procedure~\eqref{eq:estimator}, and we standardize to unit row sums later in~\eqref{equa:transform}.}

The \emph{preliminary non-standardized penalized least-squares (PNPLS) estimator} is defined as 
\begin{equation}
    \label{eq:estimator}
    \hat{\bo{\theta}}_{\fs} 
    = \rbr{\hat{A}_{\fs}, \, \hat{\bo{\theta}}_{Z, \fs}  } 
    = \argmin_{\bo{\theta} = (A,\bo{\theta}_Z) \in \aktwo{[0, 1]^{d \times r}} \times \Theta_Z}  \, 
    \cbr{
        \sum_{m=1}^q 
        \rbr{\hat{\ell}_{n,k} (\bc_m) - \ell(\bc_m ; \bo{\theta})}^2 
        + \lambda \mathcal{P}(A)
    },
\end{equation}
where $\lambda > 0$ is called the \emph{penalization parameter} and where the penalty is defined as 
$\mathcal{P}(A) = \sum_{j \in D} ( \sum_{s \in R}   a_{js}^{p} )^{1/p}$ for some \emph{penalization exponent} $p \in (0,1]$.

For $\lambda = 0$, the PNPLS estimator is equal to the ordinary least-squares estimator defined in \eqref{equa:unpen_lss}, \aktwo{provided $A$ is restricted to lie in $\Theta_A$}. 
The penalization term $\lambda \mathcal{P}(A)$ in \eqref{eq:estimator} is needed to find zero entries in the coefficient matrix $A$. We restrict our attention to penalization exponent \( p \leq  1 \). 
Geometrically, using a penalization $\mathcal{P}(A)$ with $p \leq 1$ promotes sparsity because its level sets are spiky and tend to intersect the loss contours at sparse solutions, similar to the well-known lasso ($p = 1$).
This is in contrast to the case \( p > 1 \), illustrated in Figure~\ref{Fig:spheres}. 
 In Remark~\ref{rem:update_step}, we discuss some additional arguments for the choice $p \leq  1$ related to the minimization algorithm.

\begin{figure}
    \centering
    \begin{tikzpicture}[scale=2]

    \begin{scope}[shift={(-2.5,0)}]
        \node at (0.5,2) {$p=2$};

        \draw[->] (0,0) -- (1.2,0) node[right] {$a_1$};
        \draw[->] (0,0) -- (0,1.2) node[above] {$a_2$};

        \draw[thick, red, samples=100, smooth, domain=0:90]
            plot ({cos(\x)},{sin(\x)});
    \end{scope}

    \begin{scope}[shift={(0,0)}]
        \node at (0.5,2) {$p=1$};

        \draw[->] (0,0) -- (1.2,0) node[right] {$a_1$};
        \draw[->] (0,0) -- (0,1.2) node[above] {$a_2$};

        \draw[thick, red] (1,0) -- (0,1);
    \end{scope}

    \begin{scope}[shift={(2.5,0)}]
        \node at (0.5,2) {$p=0.25$};

        \draw[->] (0,0) -- (1.2,0) node[right] {$a_1$};
        \draw[->] (0,0) -- (0,1.2) node[above] {$a_2$};

        \draw[thick, red, samples=100, smooth, domain=0:90] 
            plot ({(1 - (abs(sin(\x)))^(0.9))^(2)},
                  {(1 - (abs(cos(\x)))^(0.9))^(2)});
    \end{scope}

    \end{tikzpicture}

    \caption{\label{Fig:spheres} Bivariate contours of $(\Sigma_s a_s^{p})^{1/p} = 1$ for given values of $p$.}
\end{figure}

The minimization problem in \eqref{eq:estimator} is non-convex and is not guaranteed to give a global minimum. The parameter vector $\bo{\theta}_Z$, corresponding to the latent factors $\bZ_{\bigcdot 1}, \ldots, \bZ_{\bigcdot r}$, may be particularly difficult to estimate. 
A straightforward way to improve the PNPLS estimator is as follows. After obtaining \(\hat{\bo{\theta}}_{\fs} \), we follow-up with a two-step estimation procedure, where each step minimizes~\eqref{eq:estimator} with respect to either $\bo{\theta}_Z \in \Theta_Z$ or $A \in \Theta_A$. More precisely, 
\begin{compactenum}
    \item     let $\hat{A}_{\new }$ be the partial solution of~\eqref{eq:estimator} \wrto $A$, where $\bo{\theta}_Z$ is fixed to $\hbt_{Z, \fs}$; 
    \item  write $\hat{A}_{\new}=(\hat{a}_{js,\new})_{j \in D, s \in R}$ and standardize to ensure unit row sums,  
\begin{equation}
\label{equa:transform}
    \hat{a}_{js}= \hat{a}_{js ,  \new} \,\Big/\, \sum_{l=1,\ldots,r} \hat{a}_{jl,  \new}, \qquad j  \in D , s \in R.
\end{equation} 
This yields $\hat{A} \in [0,1]^{d \times r}$ with unit row sums;
    \item let $\hbt_{Z}$ be the partial solution of~\eqref{eq:estimator} \wrto $\bo{\theta}_Z$, \amtwo{with $\lambda=0$}, where $A$ is fixed to $\hat{A}$. 
    \end{compactenum}
The resulting estimate $\hbt = (\hat{A}, \hat{\bo{\theta}}_Z)$ is called the \emph{penalized least-squares (PLS) estimator}. \amtwo{Note that numerically, we find that $\hat{A}_{\new} \neq \hat{A}_{\fs}$. In fact, $\hat{A}_{\fs}$ is not a global minimum.}
The two steps of the estimation procedure can also be iteratively alternated until a predefined stopping criterion is met. However, as we will demonstrate in the simulation studies in Section~\ref{sec:simu_study}, performing these steps only once is typically sufficient to achieve satisfactory results in practice. 
The full estimation procedure is explained in Algorithm~\ref{alg:erstim_proced}. 

\begin{remark}
    Since \( A \) has unit row sums by definition, one can argue that the minimization problem~\eqref{eq:estimator} can be reduced to a lower-dimensional formulation where \( \Theta_A = [0,1]^{d\times(r-1)} \) by eliminating the column \( \bo{a}_{\cdot r} \) and imposing the constraint  $
a_{j1} + \dots + a_{jr-1} \leq 1, \text{for each } j \in D$. 
However, such an approach penalizes only the first \( r-1 \) columns of the matrix \( A \), leading to an asymmetric treatment of the parameters. 
\end{remark}

\begin{algorithm}[ht]
\caption{Estimation of the parameters of a mixture model for known $r$}
\label{alg:erstim_proced}
   \hspace*{\algorithmicindent} \textbf{Input:} integers $d, \, r \geq 1$, points $\bo{c}_1,\ldots, \bo{c}_q \in [0,\infty)^d$, sample $\mathcal{S}=\bX_1 , \ldots , \bX_n \in [0, \infty)^d$, sample size $n \in \mathbb{N}$, integer $k \leq n$, reals $0 <p \leq1$ and $\lambda>0$.\\
    \hspace*{\algorithmicindent} \textbf{Output:} penalized least-squares estimator $\hat{\bo{\theta}}$.
\begin{algorithmic}[1]
\State Compute the empirical stdf estimates 
$  \rbr{\hat{\ell}_{n,\, k}\rbr{\bo{c}_m} }_{m=1,\ldots,q}$ based on the sample $\mathcal{S}$. \label{step:stdf_calculus} 

\State \textbf{Preliminary non-standardized penalized least-squares estimator}: Compute the parameter estimate $\hbt_{\fs}=\rbr{\hat{A}_{\fs}, \,\hat{\bo{\theta}}_{Z, \fs}} \in \Theta_A \times \Theta_Z$ as the solution of \eqref{eq:estimator} based on empirical stdf estimates calculated in Step~\ref{step:stdf_calculus}. 



 \State \textbf{Step 1. Update the coefficient matrix}: Compute \( \hat{A}_{\new} \) as the solution of~\eqref{eq:estimator} with \( \bo{\theta}_Z \) fixed at \( \hat{\bo{\theta}}_{Z ,\fs} \) and using the starting value \( \hat{A}_{\fs} \). 

   \State \textbf{Step 2. Standardization}: Apply the transformation~\eqref{equa:transform} to \amtwo{\( \hat{A}_{\new} \)} to ensure unit row sums, \aktwo{leading to a matrix $\hat{A}$.}
   
\State \textbf{Step 3. Update the dependence parameter}: Compute \( \hat{\bo{\theta}}_{Z, \new} \) as the solution of~\eqref{eq:estimator} with \amtwo{$\lambda = 0$}, \( A \) fixed at \aktwo{\( \hat{A}  \)} and using the starting value \( \hbt_{Z , \fs} \).


\State \textbf{Standardized penalized least-squares estimator:}
Define
\begin{equation}
    \label{equa:final_estim}
    \hbt:= \rbr{\hat{A},   \hat{\bo{\theta}}_{Z} } =\rbr{\hat{A}_{ \new} ,   \hat{\bo{\theta}}_{Z, \new} } .
\end{equation}
\State \Return($\bo{\hat{\theta}}$).
\end{algorithmic}
\end{algorithm}

\begin{remark}
By Theorem~\ref{stephenson}, the extreme directions of a mixture model correspond to the set of signatures of its coefficient matrix \( A \). 
For $s \in R$, the sets  
$
\hat{J}_s = \left\{ j \in D : \hat{a}_{js} > 0  \right\}
$
thus explicitly provide estimates of the extreme directions associated with the data.
\end{remark}

\paragraph{Minimization algorithm and choice of $p$.} To solve the minimization problem~\eqref{eq:estimator}, we use the bounded limited-memory modification of the BFGS quasi-Newton method (\texttt{L-BFGS-B} algorithm in \texttt{R}), see \citet{byrd1995limited}. This iterative method has the update step 
    \begin{equation}
    \label{equa:lbfgs}
        \bo{\theta}_{m+1}= \bo{\theta}_m -\gamma_m H_m^{-1} \nabla L\rbr{\bo{\theta}_m}, \qquad m \geq 0,  
    \end{equation}
where $L$ is the loss function in~\eqref{eq:estimator}, $\gamma_m >0$ is the step-size, $H_m$ 
is a $(d\times r+v) \times (d\times r+v)$ approximate curvature positive definite matrix capturing second-order information based on the gradients observed during optimization, and $\nabla L(\btheta_m)$ is the gradient of $L$ evaluated at $\btheta_m$. At each step, the gradient is approximated using a finite-difference method. 
\begin{remark}
\label{rem:update_step}
    Another motivation for choosing the penalization exponent $p \in (0,1]$ is that the update direction in~\eqref{equa:lbfgs} is opposite to the gradient. For a fixed parameter $a_{js}$, the penalization term in~\eqref{eq:estimator} then contributes to the gradient proportionally to 
\begin{equation}
   \label{equa:dir}
T(p) =  a_{js}^{p-1} \rbr{\sum_{t \in R} a_{jt}^p } ^{1/p - 1}  .
\end{equation}  
Recall that \( p <  1 \) and thus \( p - 1 <  0 \). 
Consequently, if the true value of \( a_{js} \) is zero, then as the corresponding coefficient approaches zero at some step of the minimization algorithm L-BFGS-B, the term $T(p)$ in Equation~\eqref{equa:dir} diverges to \( +\infty \). As a result, the gradient also diverges to \( +\infty \), and since the update direction in~\eqref{equa:lbfgs} is opposite to the gradient, this leads to a solution where \( \hat{a}_{js} = 0 \).
\end{remark}


    \paragraph{Choice of tuning parameters.}
There are three tuning parameters for the estimation procedure in Algorithm~\ref{alg:erstim_proced}: the tail fraction $k/n$, the penalization exponent $p$ and the penalization parameter $\lambda$. First, the choice of the tail fraction is a classical trade-off between bias and variance; a smaller $k/n$ or larger $k/n$ introduces more variance or more bias, respectively, for the empirical stdf in~\eqref{eq:ellclassic}. Second, for $0<p_1\leq p_2 <1$, we have $T(p_1)/T(p_2)$ explodes to $\infty$ when $a_{js}$ is closer to zero, with $T$ defined in~\eqref{equa:dir}. The same reasoning after~\eqref{equa:dir} shows that a smaller penalization exponent yields stronger penalization. Different choices of $p$ will be illustrated in Section~\ref{sec:simu_study}. Finally, the penalization parameter $\lambda$ will be chosen using the $K$-fold cross validation in Algorithm~\ref{alg:cross_validation}.

\begin{algorithm}[ht]
\caption{$K$-fold cross validation score for a penalization parameter $\lambda$}
\label{alg:cross_validation}
    \hspace*{\algorithmicindent} \textbf{Input:} integers $d, \, r \geq 1$, points $\bo{c}_1,\ldots, \bo{c}_q \in [0,\infty)^d$, sample $\mathcal{S}=\bX_1 , \ldots , \bX_n \in [0, \infty)^d$, sample size $n \in \mathbb{N}$, integer $k \leq n$, reals $0 <p \leq 1$ and $\lambda>0$.\\
    \hspace*{\algorithmicindent} \textbf{Output:} cross validation score associated to the penalization parameter $\lambda$.

\begin{algorithmic}[1]
\State Split the sample $\mathcal{S}$ into $K$ (approximately) equal-sized folds $\mathcal{S}_1,\ldots,\mathcal{S}_K$. 
\For {each fold $i \in \cbr{1,\ldots,K}$}
\State \textbf{Validation set:} set $\mathcal{S}_i$ as the validation set.
\State \textbf{Training set:} set $\mathcal{S} \setminus \mathcal{S}_i$ (all other folds) as the training set.
\State Let $\hbt^{(-i)}$ be the output of Algorithm~\ref{alg:erstim_proced} with parameters $d, r$, points $\bo{c}_1,\ldots, \bo{c}_q$, sample $\mathcal{S} \setminus \mathcal{S}_i$, sample size $\lfloor n(K-1)/k \rfloor $, integer $\lfloor  k(K-1)/K \rfloor$, reals $p$ and $\lambda$.

\State Calculate the empirical stdf estimates 
 $\rbr{\hat{\ell}_{\lfloor n/K \rfloor,\, \lfloor k/K \rfloor }^{\,(i)}\rbr{\bc_m}}_{m=1,\ldots,q}$ using the validation set.
\State Evaluate the trained model on the validation set using the $i$-th score
\begin{equation*}
    \CV^{\,(-i)}(\lambda)= \sum_{m=1}^q 
\rbr{\hat{\ell}_{\lfloor n/K \rfloor,\, \lfloor k/K \rfloor }^{\,(i)}\rbr{\bc_m} - \ell\rbr{\bc_m; \bo{\hat{\theta}}^{(-i)} }}^2.
\end{equation*}
\EndFor
\State Compute the average cross validation score along all folds via 
\begin{equation*}
\CV(\lambda)= \frac{1}{K} \sum_{i=1}^{K} \CV^{\,(-i)}(\lambda).
\end{equation*}
\State \Return $\CV(\lambda)$.
\end{algorithmic}
\end{algorithm}

\aktwo{The penalized least squares estimator is consistent when we restrict the search space in the optimization problem in \eqref{eq:estimator} to a compact subset $K$ of the parameter space $\Theta_A \times \Theta_Z$ that contains the true parameter. By continuity of the objective function, a global minimum always exists on $K$. The minimizer converges in probability to the true parameter under an identifiability condition. Recall the empirical stdf $\hat{\ell}_{n,k}$ in \eqref{eq:ellclassic}.}



\begin{theorem}
    \label{thm:consistency}
    \aktwo{Let $\bX_1,\ldots,\bX_n$ be iid random vectors on $\reals^d$ with continuous marginal distribution and stdf $\ell = \ell(\cdot\,,\btheta_0) \in \cbr{\ell(\cdot\,,\btheta) : \theta \in \Theta_A \times \Theta_Z}$, following a parametrically specified mixture model as in the beginning of Section~\ref{sec:construction}. 
    Assume the points $\bo{c}_1,\ldots,\bo{c}_m \in [0, \infty)^d$ are such that the map $\btheta \mapsto \rbr{\ell(\bc_m, \btheta)}_{m=1}^q$ is continuous and injective on $\Theta_A \times \Theta_Z$. Let $K \subset \Theta_A \times \Theta_Z$ be compact and contain the true parameter $\btheta_0 = (A_0, \btheta_{Z,0})$ and define 
    \[
        \hat{\btheta}_{K}
        = \argmin_{\btheta \in K}
        \cbr{
            \sum_{m=1}^q
            \rbr{
                \hat{\ell}_{n,k} (\bc_m) - 
                \ell(\bc_m ; \bo{\theta})
            }^2 
            + \lambda \mathcal{P}(A)
        }.
    \]
    for $k = k_n \in (0, n]$ and $\lambda = \lambda_n \ge 0$.
    Then $\hat{\btheta}_{K} \to \btheta_0$ in probability as soon as $\lambda \to 0$, $k \to \infty$ and $k/n \to 0$ as $n\to\infty$.}
\end{theorem}

\aktwo{The proof of Theorem~\ref{thm:consistency} is given in Appendix~\ref{app:proof}.}

\subsection{Identifying the extreme directions}
\label{sec:EDI_algo}
In practice, the number of extreme directions $r$ is unknown. Using the same assumptions and notation 
as in Section~\ref{sec:construction}, we now estimate $\bo{\theta}$ without taking any prior knowledge of the number of columns, $r$, of $A$ into account. Our proposed method is iterative: at each step $t \geq 1$, we fix the number of columns of $A$ to $t$ and compute the standardized penalized least-squares estimator $\hat{\bo{\theta}}=\rbr{\hat{A} , \hat{\bo{\theta}}_Z} \in \Theta_A \times \Theta_Z$ 
using Algorithm~\ref{alg:erstim_proced}. We then iterate to $t + 1$ as long as all \amtwo{columns of $\hat{A}$ 
remain non-null}. The method is presented in Algorithm~\ref{alg:extr_dir_identification_and_estim_3}.

\begin{algorithm}
\caption{\label{alg:extr_dir_identification_and_estim_3} Estimation of the extreme directions associated with a data sample}
   \hspace*{\algorithmicindent} \textbf{Input:} integer $d \geq 1$, points $\bo{c}_1,\ldots, \bo{c}_q \in [0,\infty)^d$, sample $\mathcal{S}=\bX_1 , \ldots , \bX_n \in [0, \infty)^d$, sample size $n  \in \mathbb{N}$, integer $k \leq n$, reals $0 <p \leq 1$ and $\lambda>0$.\\
    \hspace*{\algorithmicindent} \textbf{Output:} extreme directions associated with the sample $\mathcal{S}$.
\begin{algorithmic}[1]
\State Set $t=1$.
\State \label{step:calcul} Let $\hat{\bo{\theta}}_t=\rbr{\hat{A} , \hat{\theta}_Z}$ be the output of Algorithm~\ref{alg:erstim_proced} with parameters $d,t$, points $\bc_1,\ldots, \bc_q$, sample $\mathcal{S}$, sample size $n$, integer $k\leq n$, reals $p$ and $\lambda$.
\State \label{step:sign} Compute the signatures $\hat{J}_s = \cbr{j \in D: \, \hat{a}_{js} >0 }$ for $s = 1,\ldots,t$ and let $\hat{\mathcal{J}} = \{\hat{J}_1,\ldots,\hat{J}_t\}$.
\While {$\varnothing \notin \hat{\mathcal{J}}$} 
\State Iterate $t \leftarrow t+1$.
\State  Repeat Steps~\ref{step:calcul} and~\ref{step:sign}.
\EndWhile
\State Remove $\cbr{\varnothing}$ from $\hat{\mathcal{J}}$, that is, update $\hat{\mathcal{J}} \leftarrow \hat{\mathcal{J}} \setminus \cbr{\varnothing}$.

\State \Return $\hat{\mathcal{J}}$.
\end{algorithmic}
\end{algorithm}

The DAMEX algorithm introduced by \citet[Algorithm~1]{goix2017JMVA} can also be used to identify extreme directions. It relies on three key parameters: a threshold beyond which data are considered extreme, analogous to the tail fraction $k/n$ used in Algorithm~\ref{alg:erstim_proced}; a tolerance parameter that heuristically defines the region associated with an extreme direction; and a mass threshold, which determines whether a subset $J \subseteq V$ with sufficiently large mass is considered an extreme direction. As explained in \citet{goix2017JMVA}, in a supervised setting, these parameters are selected via cross-validation. As we will detail in the simulation study in Section~\ref{sec:simu_study}, this is also the case for Algorithm~\ref{alg:erstim_proced}: the choice of the tail fraction involves a trade-off between bias and variance. In contrast, the choice of the penalization exponent $p$ is less critical, provided that the penalization parameter $\lambda$ is appropriately tuned—this being achieved, as outlined in Algorithm~\ref{alg:cross_validation}, through $K$-fold cross validation.


Our procedure is tailored to max-stable mixture models, \amtwo{motivated by \citet[Theorem~4.8]{mourahib2024multivariate}, who show that every max-stable distribution with unit-\FR~margins can be represented as a max-stable mixture model. In practice, the factors $\bZ_{\bigcdot 1}, \ldots, \bZ_{\bigcdot t}$ are assumed to follow a parametric model, ensuring that each factor has a single extreme direction $D$. While this could be seen as a limitation of our method, it has little impact on the identification of extreme directions.  Indeed, for max-stable mixture models, the set of extreme directions is determined solely by the matrix $A$, and does not depend on the specific parametric form imposed on the factors. This claim will be supported by simulation results in Section~\ref{sec:simu_study}.}

\section{Simulation study}
\label{sec:simu_study}
In this section, we conduct a simulation study to evaluate the estimation procedure introduced in Section~\ref{sec:estimation_mixture_model}. We begin by outlining the setup used throughout the simulations in Section~\ref{sec:preliminaries}. First we consider the setting where the number of extreme directions is known, as described in Algorithm~\ref{alg:erstim_proced} and detailed in Section~\ref{sec:known_num_ext_dir}. We then turn to the more general case in which the number of extreme directions is unknown, corresponding to Algorithm~\ref{alg:extr_dir_identification_and_estim_3}, and discussed in Section~\ref{sec:simu_uknown}.

\subsection{Preliminaries}
\label{sec:preliminaries}

\paragraph{Coefficient matrix.} Throughout the simulation study, we fix the dimension $d=4$ and the coefficient matrix 
\begin{equation}
    \label{equa:matrix_A2}
    A=
   \begin{pmatrix}
        1/3 & 1/3 & 1/3 & 0 \\
        1/2 & 0 & 0 & 1/2 \\
        0 & 3/4 & 1/4 & 0 \\
        0 & 1/2 & 0 & 1/2 
    \end{pmatrix}.
\end{equation}

\paragraph{Domain of attraction.}
We simulate samples $\bX_1,\ldots, \bX_n$ of size $n$ in the domain of attraction of a mixture model $\bM$ as defined in~\eqref{mixturemodel}. To do so, we simply perturb $\bM$ by adding a lighter-tailed noise. More precisely, 
\begin{equation}
\label{equa:simu}
    \bX_{i}= \bM_i + \bo{N}_i, \qquad i=1,\ldots,n,
\end{equation}
where for $i=1,\ldots,n$, $\bM_i \in \reals^d$ follows a mixture model and $\bo{N}_i \in \reals^d$ has independent centered Gaussian margins with variance $\sigma^2>0$. Algorithm~\ref{alg:N_generate_Mix_log} \amtwo{in Appendix~\ref{app:algo}} outlines the procedure for simulating from a mixture model, relying on the generation of the max-stable random vectors $\bZ_{\bigcdot 1 },\ldots, \bZ_{\bigcdot r}$. This is feasible for most popular models; see \citet[Algorithm~1]{dombry2016exact} 
or the \texttt{R} package \texttt{mev} \citep{belzile2024package}. 
Furthermore, code for Algorithm~\ref{alg:N_generate_Mix_log} is freely accessible for both the mixture logistic model and the mixture \HR~model.\footnote{\url{https://github.com/AnasMourahib/Penalized_least-squares_estimator/blob/main/Functions/Simulation_mixture_model.R}.}

\paragraph{Starting points in the minimization algorithm.}
To calculate the PNPLS estimator, 
we need to provide good starting values for $\bo{\theta}=(A , \bo{\theta}_Z)$. Suppose that the number of extreme directions is fixed to $t$, and therefore $A \in [0,1]^{d \times t}$. Consider a $d$-variate sample $\bX_1,\ldots, \bX_n$ and write $\bX_i= (X_{i1},\ldots, X_{id})$, for $i=1,\ldots,n$. We determine a starting value for $A$ by applying the $k$-means 
clustering algorithm 
with $t$ clusters to the unit-Pareto transformed data,
\[
\left( \frac{n}{n + 1 - R_{i1}} , \ldots , \frac{n}{n+1 - R_{id}} \right), \qquad i=1,\ldots, n ,
\]
where for $i=1,\ldots, n$ and $j=1,\ldots, d$, $R_{ij}$ is the rank of $X_{ij}$ among $X_{1j},\ldots, X_{nj}$. We retain only the top $10\%$ of observations with the highest sum of transformed values. 

Starting values for the dependence parameters $\bo{\theta}_Z$ will be assigned randomly, depending on the selected model. As in Section~\ref{sec:estimation_mixture_model}, recall that for simplicity, we suppose that the dependence parameters across the columns of $A$ are the same.

\paragraph{Grid points $\bo{c}_1,\ldots, \bo{c}_q$.} Both the parametric and the empirical stdf are evaluated in $\bo{c}_1,\ldots, \bo{c}_q$ during the minimization~\eqref{eq:estimator}. 
\aktwo{A sufficiently large number of points \(q\) ensures reliable estimation, and does not lead to computational difficulties even with a larger dimension: the evaluation of the
empirical stable tail dependence function is fast, as it does not enter
the optimization problem.} We use the \texttt{R} package \texttt{tailDepFun} \citep{kiriliouk2016vignette}, we select points that can be generated in a $4$-dimensional space:
\begin{itemize}
    \item \textbf{for the mixture logistic fit:} using values $\cbr{1/4, \, 1/3 ,\,  1/2, \, 3/4 , \, 1}$  \aktwo{where each $\bm{c}_m$ has either two or three non-zero components}. This gives $q=650$ points;
    \item \textbf{for the mixture \HR~fit:} using values $\cbr{1/6, \, 1/8 ,\,  1/4, \, 1/3 , \, 2/3 , \, 3/4 , \, 1}$, \aktwo{where each $\bm{c}_m$ has two non-zero components}. This gives $q=384$ points.
\end{itemize} 
\amtwo{
For the mixture \HR~model, the grid includes more points with only two non-zero components than does the mixture logistic model. This choice is motivated by the fact that, in the former,
the dependence coefficients \(\Gamma_{i,j}\) of the variogram matrix must be estimated for each pair \((i,j)\), while for the latter, the dependence coefficient $\alpha$ is the same for all pairs.} 

\paragraph{Performance assessment.} We simulate $N = 100$ samples as in~\eqref{equa:simu} and for each sample, we obtain the PLS estimator $\hbt^{(l)}= (\hat{A}^{(l)} ,   \hat{\bo{\theta}}_{Z}^{(l)} ) $, $l = 1,\ldots,N$, using Algorithm~\ref{alg:erstim_proced}. \aktwo{We focus on the mixture logistic model, for which $\bo{\theta}_Z  \in (0,1)$, and the mixture \HR{} model, for which $\bo{\theta}_Z  \in (0,\infty)^{d(d-1)/2}$.} 

First, we assess the performance of our estimator in terms of the identification of extreme directions. We use a score, denoted ``\( \EDS \)'', which represents the Jaccard distance and computes the fraction of misidentified extreme directions, 
\begin{align}
\label{equa:ED-S}
\EDS
=\frac{1}{N} \sum_{l=1}^N  \D(\hat{\mathcal{J}}_l, \mathcal{J} ),   
\qquad 
\D(\hat{\mathcal{J}}_l, \mathcal{J} ) 
= 1 - \frac{\lvert \hat{\mathcal{J}}_l \cap \mathcal{J} \rvert}{\lvert \hat{\mathcal{J}}_l \cup \mathcal{J} \rvert}
= \frac{\lvert \hat{\mathcal{J}}_l \symdif \mathcal{J} \rvert}{\lvert \hat{\mathcal{J}}_l \cup \mathcal{J} \rvert}
\end{align}
where $\mathcal{J}$ and $\hat{\mathcal{J}}_l$ denote the signatures of $A$ and $\hat{A}^{(l)}$, respectively, as defined in~\eqref{eq:Jk}, \amtwo{and $\Delta$ denotes the symmetric difference between two sets as defined in Section~\ref{sec:introduction}.}
Second, we assess the performance of our estimator by measuring its proximity to the true parameter 
\( \bo{\theta} = \rbr{A , \bo{\theta}_Z } \) of the mixture model. In the following, for $l = 1, \ldots, N$, we reorder the columns of $\hat{A}^{(l)}$ so as to minimize the \aktwo{difference 
$\sum_{j \in D} \sum_{s \in R } (\hat{a}_{ js}^{(l)} - a_{js})^2.$
}

For simplicity, we continue to denote the resulting matrix by $\hat{A}^{(l)}$. To construct a robust performance metric and ensure that the marginal components of our estimator are equally scaled, we mitigate the disproportionate influence of their magnitudes. This is achieved by normalizing each coefficient of $\hbt$ using the maximum value computed over the \( N \) estimates corresponding to that component. 
More precisely, let \( \hat{m}_A = (\hat{m}_{A, js})_{j \in D , \, s  \in R}  \) denote the $(d \times r)$ matrix where the \((j,s)\)-th entry represents the maximum value calculated based on the coefficients \( \hat{a}_{js}^{(l)} \) obtained from the matrices $\hat{A}^{(l)}$, for $l = 1,\ldots,N$. Similarly, recall from Section~\ref{sec:estimation_mixture_model} the dimension $v$ of the dependence parameter vector $\hbt_Z$. Let $ \hat{\bo{m}}_{Z} = (\hat{m}_{Z , t})_{t=1,\ldots, v}$ denote the $v$-dimensional vector where the $t$-th entry represents the maximum value calculated based on the coefficients \( \hat{\theta}_{t}^{(l)} \) obtained from the vectors $\hbt_Z^{(l)}$, for $l = 1, \ldots , N$.  
We define a standardized version of the mean-squared error as
\begin{align} 
\label{equa:RMSE2}
    \RRMSE
    &=   \frac{1}{N}  \sum_{l=1}^{N} \sbr{ \diffe_{m_A}\rbr{\hat{A}^{(l)} , \, A} + \diffe_{m_Z} \rbr{\hat{\bo{\theta}}_{Z}^{(l)} , \bo{\theta}_Z  } } , 
    \quad  \text{ where} \\
    \diffe_{A}\rbr{\hat{A}^{(l)} , \, A}
    &= \sum_{j \in D} \sum_{s \in R } \frac{\rbr{\hat{a}_{ js}^{(l)} - a_{js}}^2}{\hat{m}_{A, js}^2},  
    \quad 
    \diffe_{Z} \rbr{\bo{\hat{\theta}}_{Z}^{(l)} , \bo{\theta}_Z  } 
    = \sum_{t= 1}^v \frac{\rbr{\hat{\theta}_{Z ,t}^{(l)}  - \theta_{Z,t} }^2}{\hat{m}^2_{Z,t}}. \label{equa;scores}
\end{align} 
Note that the maximum of a coefficient across \( N \) estimates is zero if and only if all estimates of that coefficient are equal to zero. This occurs only for the null coefficients of \( A \). In such cases, by convention, we define \( 0/0 = 0 \).

\subsection{Known number of columns}
\label{sec:known_num_ext_dir}

We consider the mixture logistic model in Example~\ref{example:mix_log} and the mixture \HR~model in Example~\ref{example:mix_HR}, both with coefficient matrix \( A \) in~\eqref{equa:matrix_A2} and
\begin{compactenum}
    \item \label{item:mix_log} dependence parameter \( \alpha := \alpha_1 = \ldots = \alpha_4 = 0.25 \) for the mixture logistic model,
    \item \label{item:mix_HR} variogram matrix $\Gamma := \Gamma^{(1)} = \ldots = \Gamma^{(4)}$ with all off-diagonal elements equal to $1$ for the mixture \HR{} model.
\end{compactenum}
 Note that for the mixture \HR~model, we do not suppose that all the off-diagonal elements of the variogram matrix are equal in the estimation procedure, meaning that $\Theta_Z \subseteq (0,\infty)^6$ in~\eqref{eq:estimator}. 
 
 In this section, we suppose that the number of extreme directions, $r=4$, is known.  We generate $N=100$ samples of size $n \in \cbr{1\,000 ,\, 2\,000 ,\, 3\,000}$ in the domain of attraction of the mixture logistic model and of the mixture \HR~model as in~\eqref{equa:simu}, both with a noise variance set to $\sigma^2 = 9$.

\paragraph{Effect of the penalization exponent.}
We start by studying the effect of the penalization exponent $p$ used in the minimization problem~\eqref{eq:estimator} on the scores $\EDS$ in~\eqref{equa:ED-S} and $\RRMSE$ in~\eqref{equa:RMSE2}. For the tail fraction, we set $k/n =n^c/n$ with $c=\cbr{0.4,0.5,0.6}$. With sample sizes $n \in \{1\,000, \, 2\,000, \, 3\,000\}$, 
this corresponds approximately to $k/n \in \cbr{0.01 , 0.02 , 0.05}$.  Recall that we choose the penalization parameter $\lambda$ in~\eqref{eq:estimator} based on a $10$-fold cross validation (Algorithm~\ref{alg:cross_validation}) over a well chosen grid. We use an adaptive penalization parameter grid, meaning that for each value of the penalization exponent \( p \), we have a distinct grid of penalization parameters $\lambda$. 

\begin{figure}[ht]
      \centering
    \begin{minipage}[t]{0.325\textwidth}
        \includegraphics[width=1\textwidth]{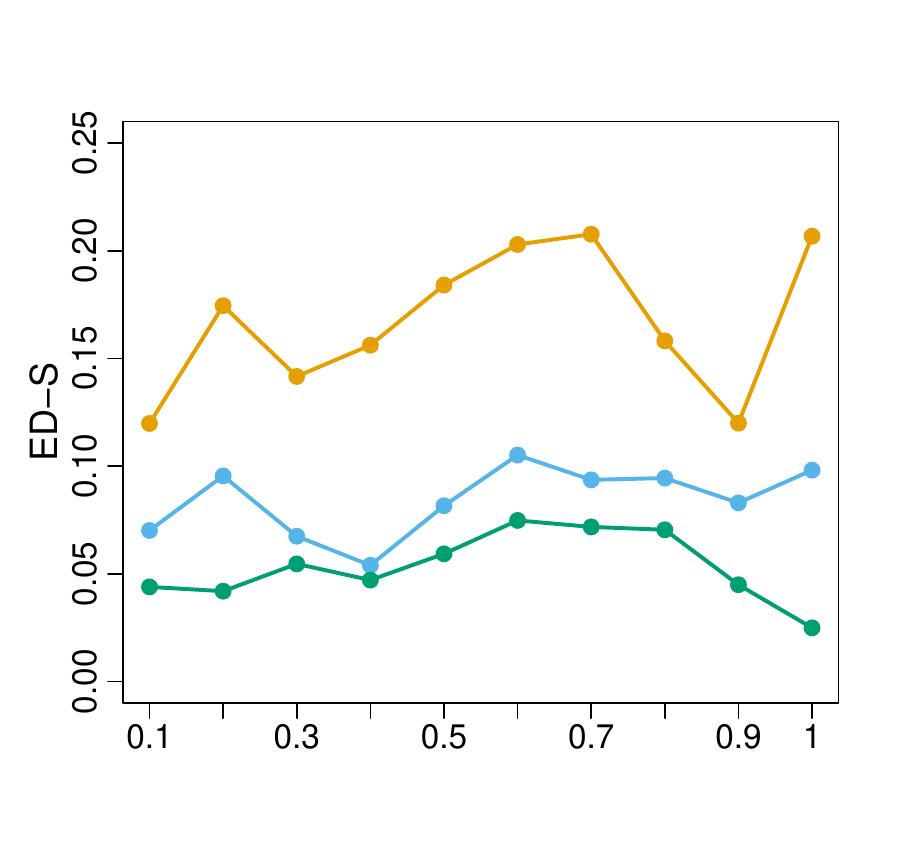} 
    \end{minipage}%
    \begin{minipage}[t]{0.325\textwidth}
        \includegraphics[width=1\textwidth]{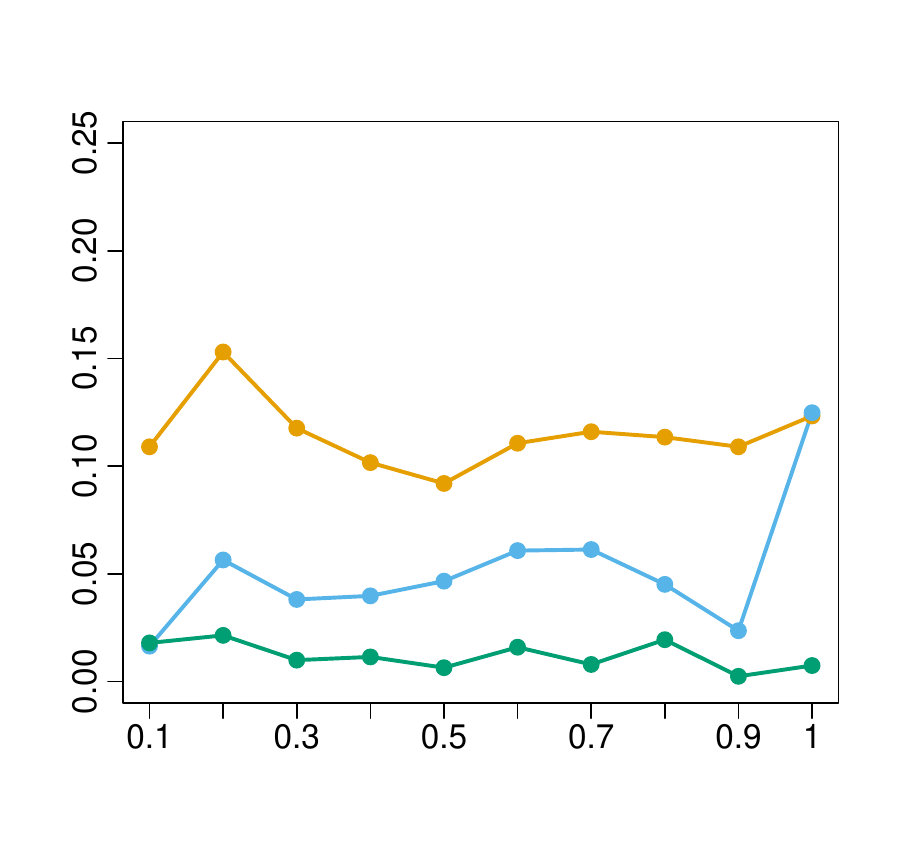} 
    \end{minipage}
     \begin{minipage}[t]{0.325\textwidth}
        \includegraphics[width=1\textwidth]{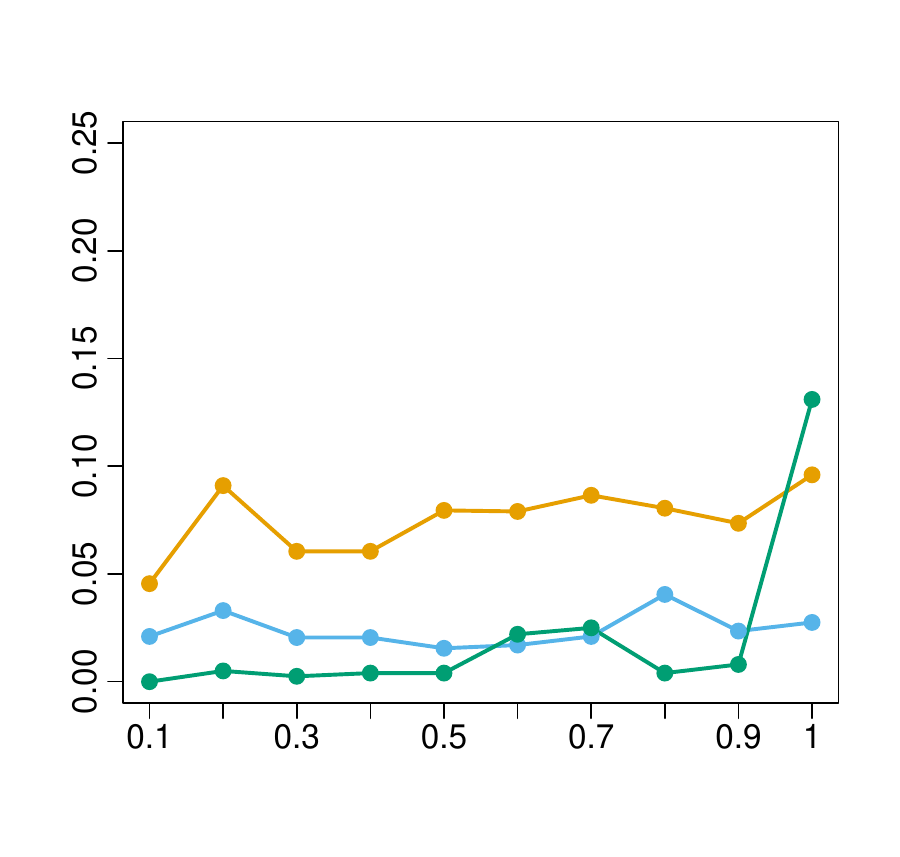} 
    \end{minipage}
    \\ 
    
    \begin{minipage}[t]{0.325\textwidth}
        \includegraphics[width=1\textwidth]{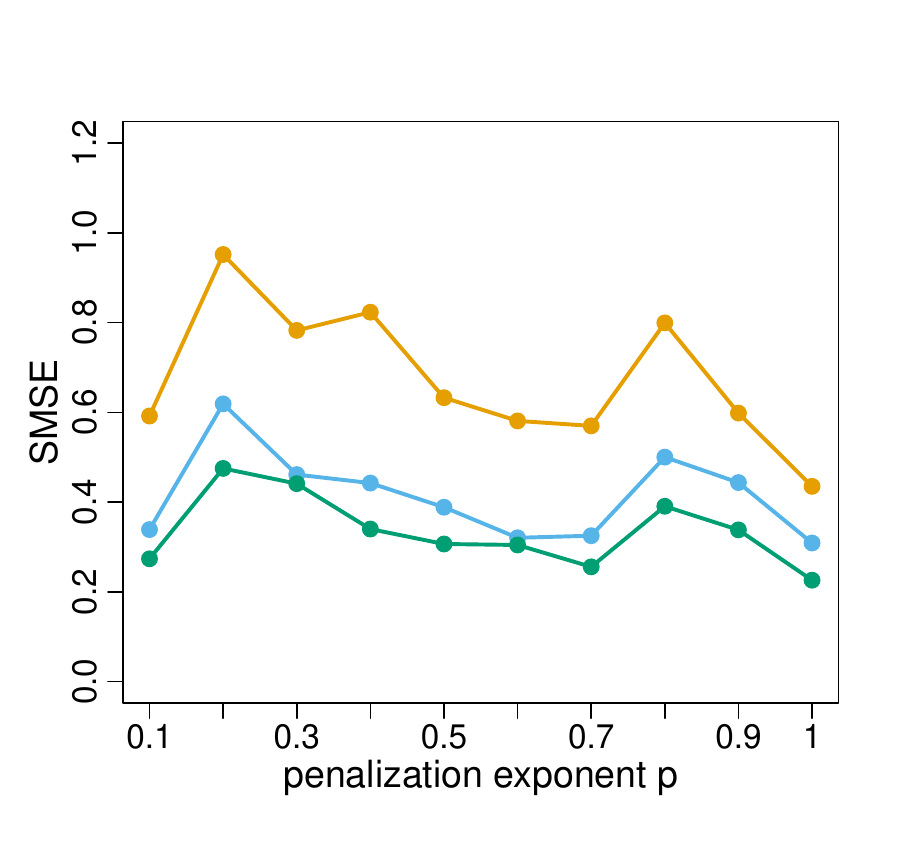} 
    \end{minipage}%
    \begin{minipage}[t]{0.325\textwidth}
        \includegraphics[width=1\textwidth]{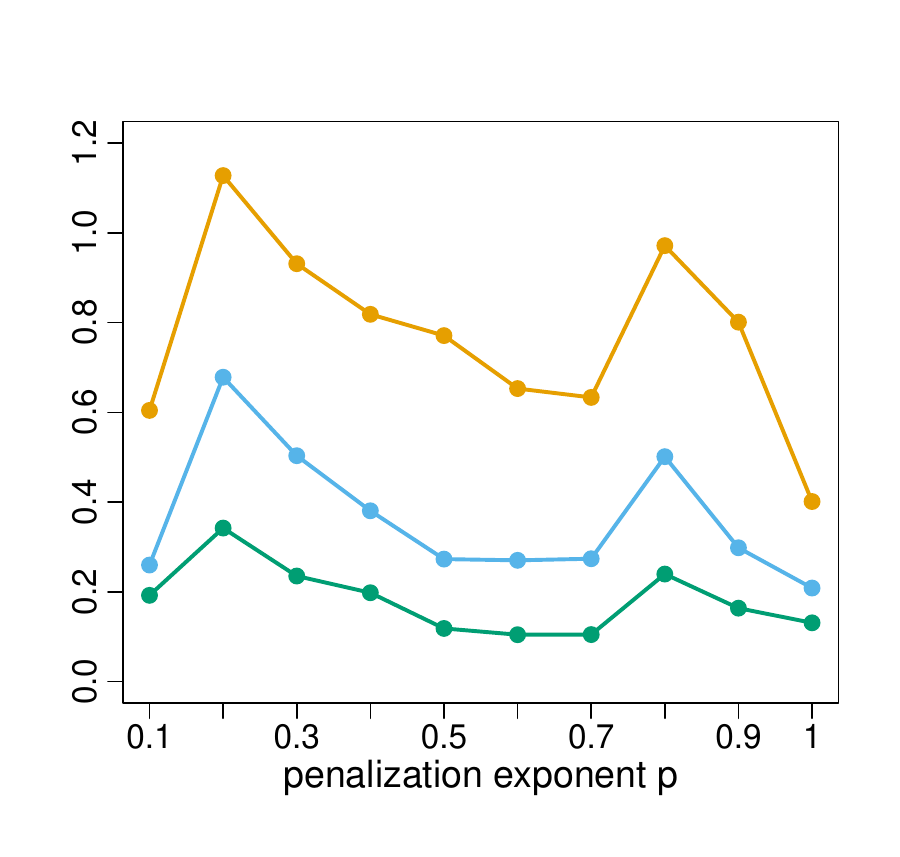} 
    \end{minipage}%
     \begin{minipage}[t]{0.325\textwidth}
        \includegraphics[width=1\textwidth]{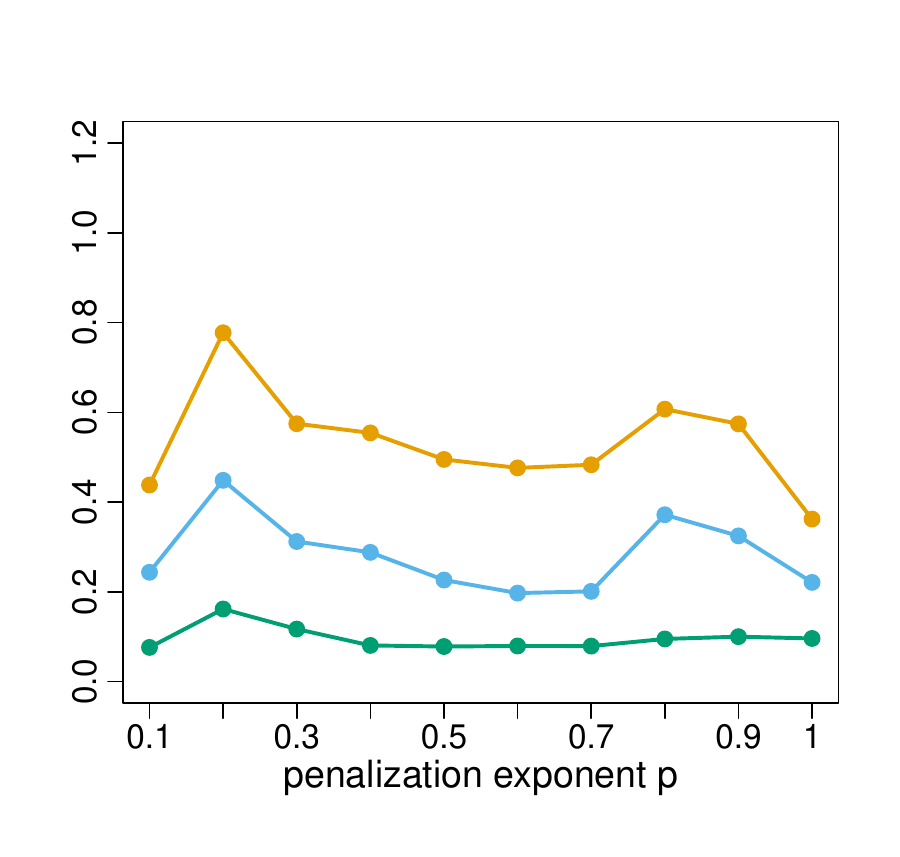} 
    \end{minipage}%

 \caption{\label{fig:effect_p} 
       Scores $\EDS$ (top) and $\RRMSE$ (bottom) defined in~\eqref{equa:ED-S} and~\eqref{equa:RMSE2}, respectively, as functions of the penalization exponent $p$, 
       for the mixture logistic model perturbed as in~\eqref{equa:simu} and sample sizes $n = 1 \, 000$ (left), $n = 2 \, 000$ (center), and $n = 3\, 000$ (right). Curves correspond to varying tail fractions, with approximately $k/n = 0.01$ (orange), $k/n = 0.02$ (blue), and $k/n = 0.05$ (green).}
\end{figure}

Figure~\ref{fig:effect_p} displays the scores $\EDS$ and $\RRMSE$ as functions of the penalization exponent $p$. Results are shown for the mixture logistic fit. Qualitatively similar results arise under the mixture \HR~fit. The effect of $p$ is not important as long as the penalization parameter $\lambda$ in~\eqref{eq:estimator} is well chosen. 
In fact, for $0 < p_1 < p_2 < 1$ and a \aktwo{coefficient $a$ of $A$}, the term $T$ in~\eqref{equa:dir}, which governs the opposite direction step in the minimization algorithm, satisfies $T(p_1)/T(p_2) \to \infty$, whenever the parameter $a \to 0$. It follows that the penalization in~\eqref{eq:estimator} is stronger with smaller penalization exponent. This explains the use of an adaptive grid of penalization parameters, where the values of the grid associated with $p_1$ are smaller than the ones used for $p_2$. 

\paragraph{Effect of the tail fraction.} We study the effect of the tail fraction $k/n$ used in~\eqref{eq:estimator} on the scores $\EDS$ in~\eqref{equa:ED-S} and $\RRMSE$ in~\eqref{equa:RMSE2}. Figure~\ref{fig:effect_p} shows that the value of the penalization exponent $p$ is not important as long as the penalization parameter $\lambda$ is well chosen using $10$-fold cross validation. Therefore, we set $p=0.4$ and perform a $10$-fold cross validation to choose $\lambda$, where for each value $p$, we use an adapted grid. 

\begin{figure}
      \centering
    \begin{minipage}[t]{0.45\textwidth}
        \includegraphics[width=1\textwidth]{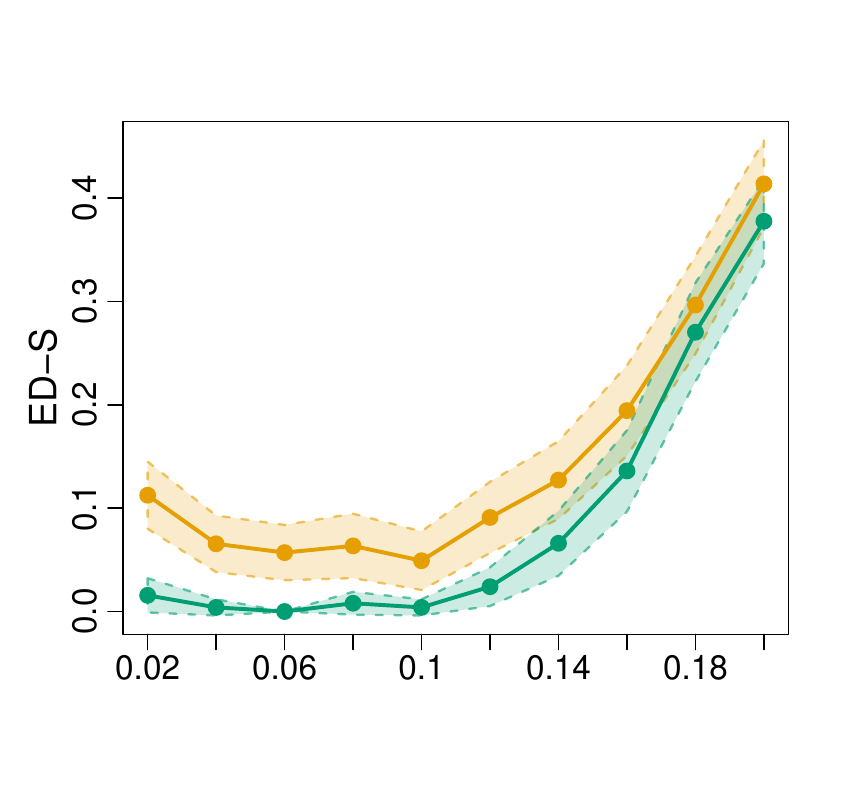} 
    \end{minipage}%
    \begin{minipage}[t]{0.45\textwidth}
        \includegraphics[width=1\textwidth]{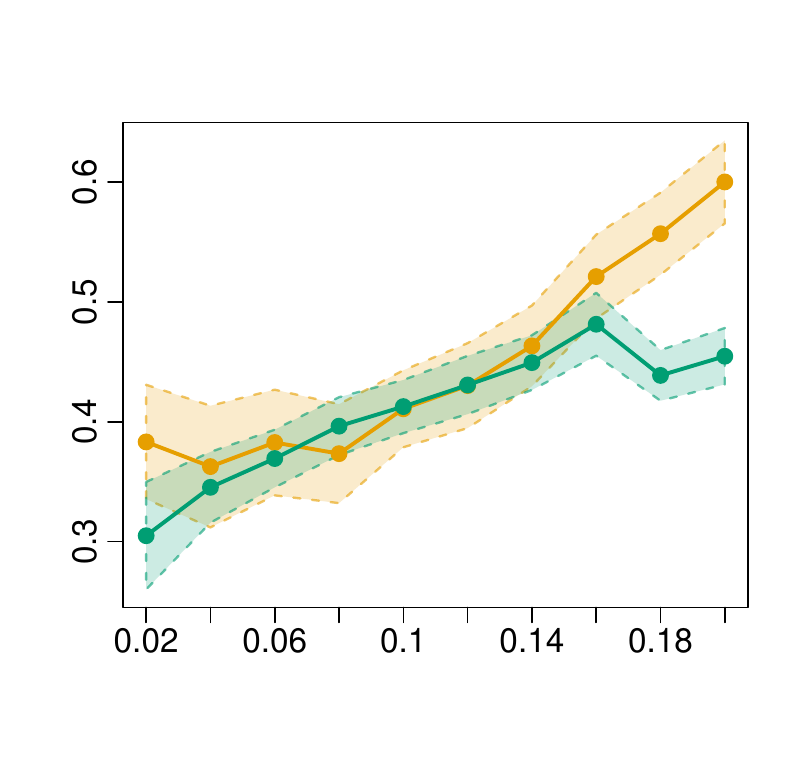} 
    \end{minipage}
    \\
    \begin{minipage}[t]{0.45\textwidth}
        \includegraphics[width=1\textwidth]{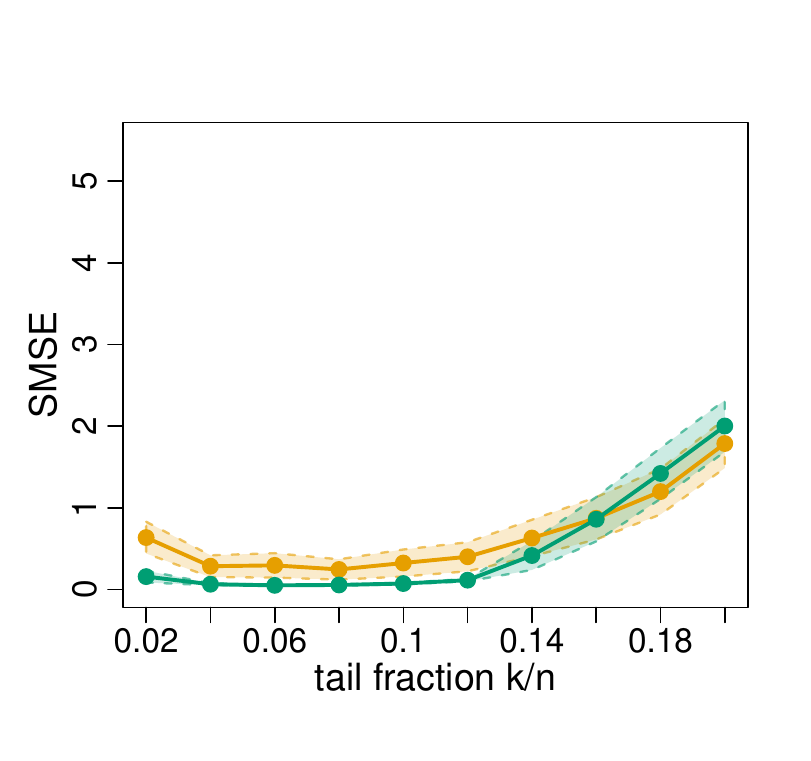} 
    \end{minipage}%
    \begin{minipage}[t]{0.45\textwidth}
        \includegraphics[width=1\textwidth]{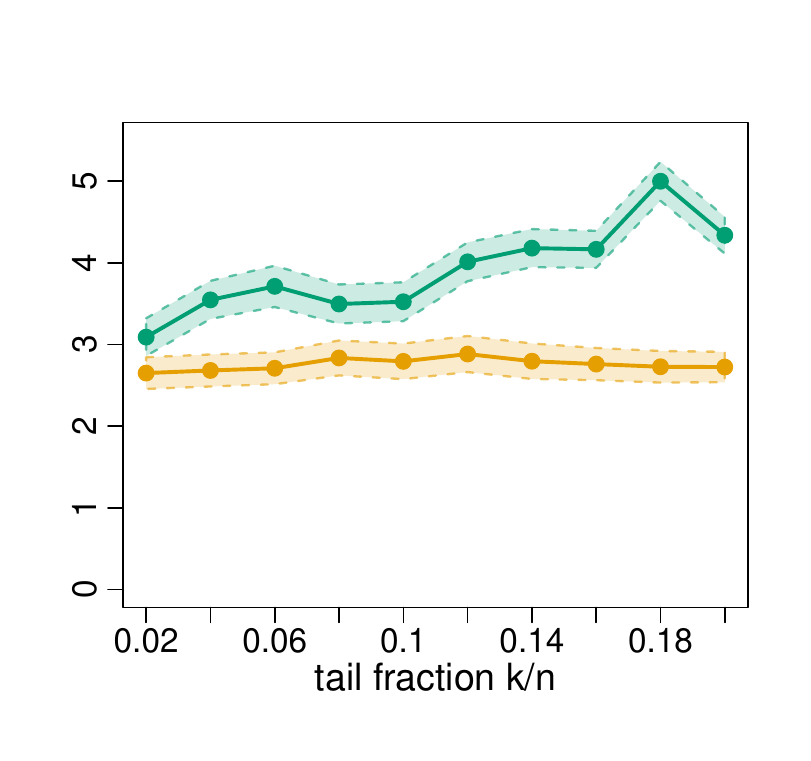} 
    \end{minipage}
\caption{\label{Figure:SMSE}
\amtwo{Scores $\EDS$ (top) and $\RRMSE$ (bottom) defined in Equations~(4.3) and~(4.4) from the paper, respectively, as functions of the tail fraction $k/n$. The shaded regions represent pointwise $95\%$ confidence bands, constructed from the empirical $2.5\%$ and $97.5\%$ quantiles of the two scores. Results are shown for the mixture logistic model (left) and the mixture \HR~model (right), both perturbed by the addition of independent multivariate Gaussian noise with independent margins. The penalization exponent is fixed at $p = 0.4$. The curves correspond to sample sizes $n = 1\,000$ (orange),  and $n = 3\,000$ (green).}}
\end{figure}

Figure~\ref{Figure:SMSE} illustrates the behavior of the scores $\EDS$ and $\RRMSE$ as functions of the tail fraction $k/n$ for a mixture logistic and a mixture \HR~fit, respectively.

\amtwo{For the mixture logistic model, both scores tend to improve toward the middle of the tail fraction grid. Additionally, the $95\%$ confidence bands for both scores become narrower in this region.}
This pattern reflects a bias–variance trade-off: a smaller tail fraction $k/n$ results in higher variance but lower bias, as fewer points are treated as extreme observations. In contrast, a larger $k/n$ leads to lower variance but increased bias, since more points—possibly not truly extreme—are included in the estimation. We also observe that as the sample size increases, the optimal choice of the tail fraction $k/n$—that is, the optimal proportion of points considered as extreme—tends to decrease. This is intuitive: with a larger sample size, a smaller proportion of extreme observations is sufficient to achieve accurate estimation.

 The score $\EDS$ for the mixture \HR~fit behaves similarly to that of the mixture logistic fit. In contrast, the score $\RRMSE$ does not reflect the bias-variance trade-off effect observed for the mixture logistic fit. We can also see that this score is higher for the mixture \HR~model. This makes sense, \aktwo{since the dependence parameter of the mixture \HR~model is of dimension $v = 6$, while the dependence parameter of the mixture logistic model is of dimension $v=1$}. This makes the estimation of the dependence parameter for the mixture \HR~model more complicated.

\begin{figure}[ht]
      \centering
            \includegraphics[width=0.6\textwidth]{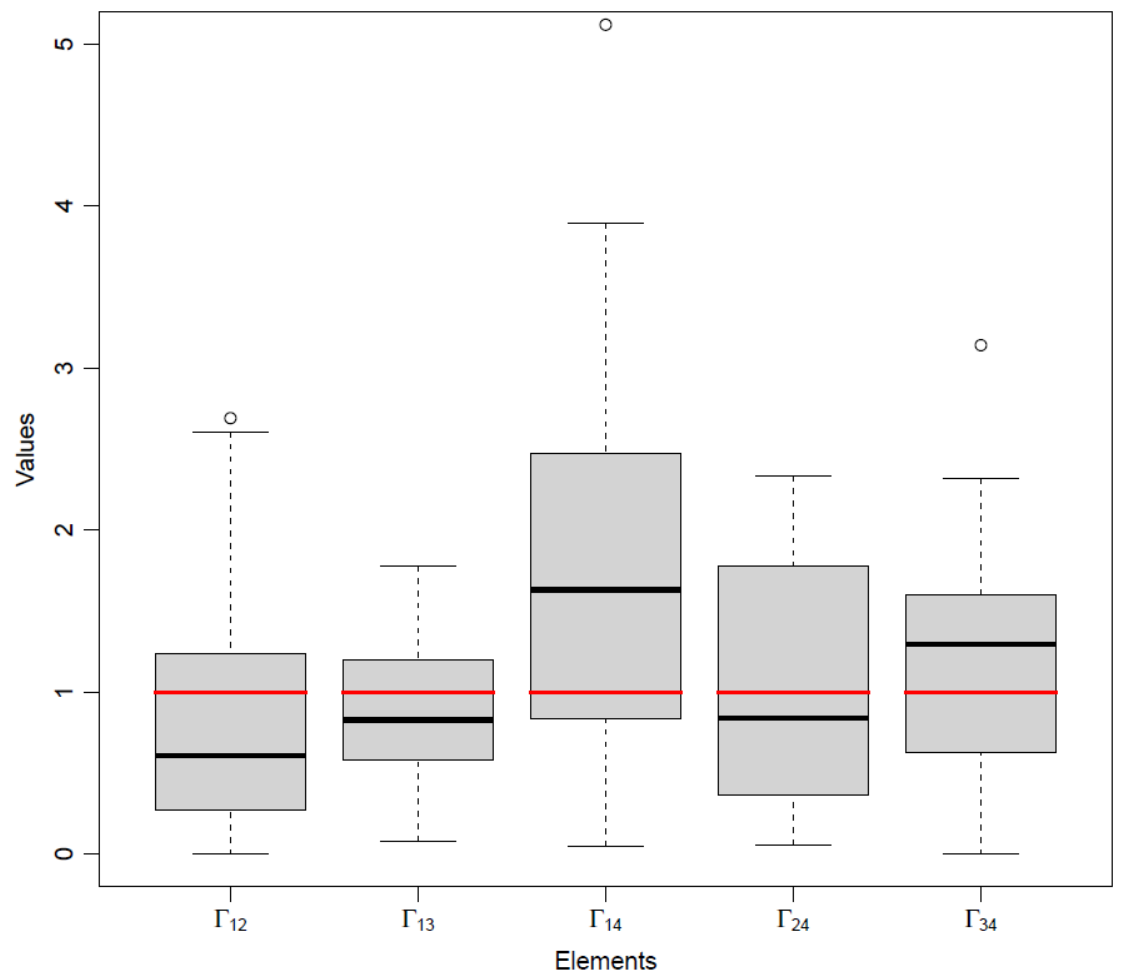} 
\caption{\label{fig:boxplots}
Boxplots of the estimated coefficients $\Gamma_{ij}$ from the variogram matrix $\Gamma$ for all pairs $(i,j)$ included in some signature $J$ of $A$. The true value $1$ for each coefficient is presented with a horizontal red line. The tail fraction  $k/n$ is set to $0.06$, the penalization exponent to $p= 0.4$ and the sample size to $n= 1\,000$.}
\end{figure}

We now discuss the estimation of the variogram matrix $\Gamma$ of the mixture \HR~model. 
Figure~\ref{fig:boxplots} shows the boxplots of the  upper off-diagonal entries of $\Gamma$, except for the entry $\Gamma_{23}$; since no signature of the matrix $A$ in~\eqref{equa:matrix_A2} contains the pair $(2,3)$, the coefficient $\Gamma_{23}$ is not identifiable.
More precisely, $\Gamma_{23}$ does not contribute to the stdf in~\eqref{stdfgeneralmodel} associated to the corresponding mixture model.


\subsection{Unknown number of columns}
\label{sec:simu_uknown}

\begin{figure}
      \centering
    \begin{minipage}[t]{0.45\textwidth}
        \includegraphics[width=1\textwidth]{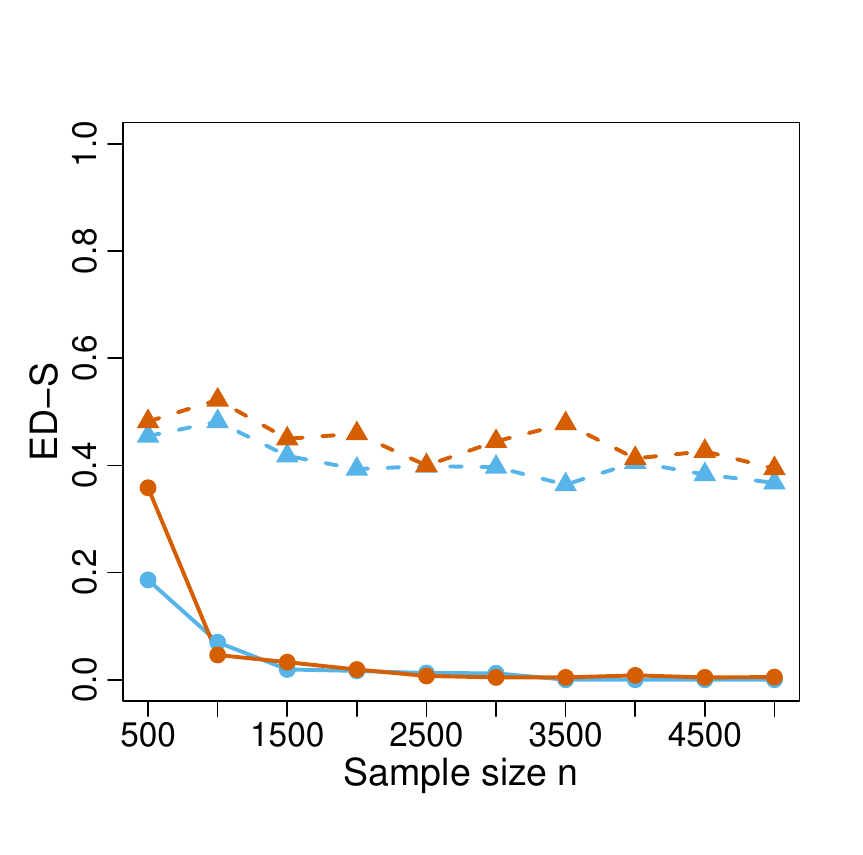} 
    \end{minipage}%
    \begin{minipage}[t]{0.45\textwidth}
        \includegraphics[width=1\textwidth]{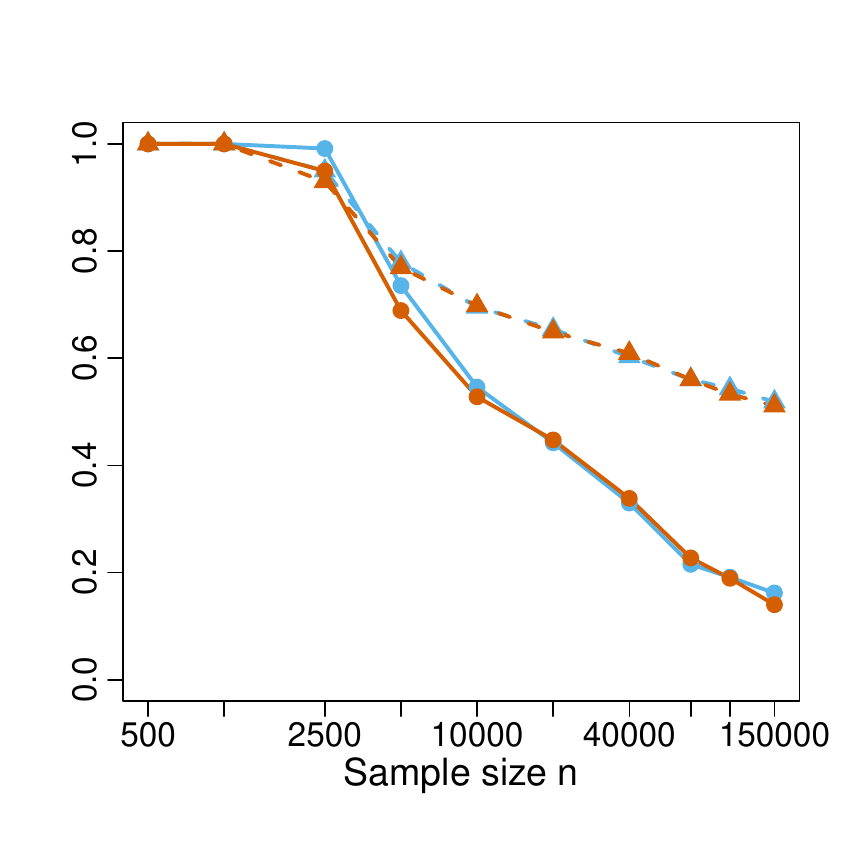} 
    \end{minipage}
 \caption{\label{fig:effect_N}
\amtwo{The score $\EDS$, defined in~\eqref{equa:ED-S}, for simulations from the mixture logistic model (blue) and the mixture \HR~model (dark orange). Solid lines correspond to data perturbed as in~\eqref{equa:simu}, while dashed lines represent data perturbed as in~\eqref{equa:simu2}. Results on the left are obtained using Algorithm~\ref{alg:extr_dir_identification_and_estim_3}, with a mixture logistic fit using $p = 0.4$ and a tail fraction of $k/n = 0.08$. Results on the right are obtained using the DAMEX algorithm from~\citet[Algorithm~1]{goix2017JMVA}, with tuning parameters $k = n^{1/2}$ and $\epsilon = 0.01$, as recommended in~\citet[Remark~7]{goix2017JMVA}. Note that the sample size used for the DAMEX algorithm reaches the order of $10^5$, whereas the sample size used for Algorithm~\ref{alg:extr_dir_identification_and_estim_3} is of order $10^3$.}
}  
\end{figure}

We now suppose that the number of extreme directions is unknown and identify the extreme directions using Algorithm~\ref{alg:extr_dir_identification_and_estim_3}. Based on Figure~\ref{fig:effect_p}, the choice of the penalization parameter $p$ is not important as long as the penalization parameter $\lambda$ is well chosen using $K$-fold cross validation as explained in Algorithm~\ref{alg:cross_validation}. Therefore, we simply choose $p=0.4$ and use a $10$-fold cross validation procedure to choose $ \lambda$ from the grid associated with $p=0.4$. Based on Figure~\ref{Figure:SMSE}, we choose $k/n \in \cbr{0.04 , 0.06 , 0.08}$.
\amtwo{We simulate independent samples $\bo{L}_1, \ldots, \bo{L}_n \in \reals^d$ from the mixture logistic model described in Point~\ref{item:mix_log}, and samples $\bo{Y}_1, \ldots, \bo{Y}_n \in \reals^d$ from the mixture \HR~model introduced in Point~\ref{item:mix_HR}.
As a first step, both models are perturbed by adding independent centered Gaussian noise, as defined in~\eqref{equa:simu}.
In a second, more challenging experiment, we perturb both models by adding positively associated Gaussian noise with \FR~margins. More precisely,
\begin{equation}
\label{equa:simu2}
\bX_i = \bo{M}_i + \bo{U}_i, \qquad i = 1, \ldots, n,
\end{equation}
where, for each $i = 1, \ldots, n$, $\bo{M}_i$ denotes either $\bo{L}_i$ or $\bo{Y}_i$, and $\bU_i \in \mathbb{R}^d$  is a random vector with unit \FR~margins and a Gaussian copula exhibiting positive dependence, with correlation matrix  $R$ such that  $R_{st}$ is chosen randomly from $[0, 0.1]$ for two pairs  $s \neq t$ from $\{1, \ldots, d\}$.}
We apply Algorithm~\ref{alg:extr_dir_identification_and_estim_3} with the following two settings: 
\begin{compactenum}
    \item \label{item:per_mi_log} \textbf{Correct model specification.} Fit a mixture logistic model to $\bX_1,\ldots, \bX_n$ and a mixture \HR~model to $\bY_1,\ldots, \bY_n$;
    \item \label{item:mix_HR2} \textbf{Misspecified model.} Fit a mixture logistic model to $\bY_1,\ldots, \bY_n$ and a mixture \HR~model to $\bX_1,\ldots, \bX_n$.
\end{compactenum}
\amtwo{We study the effect of the sample size $n$ on the score $\EDS$ defined in~\eqref{equa:ED-S}. Figure~\ref{fig:effect_N} shows the results obtained when applying both Algorithm~\ref{alg:extr_dir_identification_and_estim_3} with a mixture logistic fit and the DAMEX algorithm from~\citet[Algorithm~1]{goix2017JMVA}. We start by presenting the results for Algorithm~\ref{alg:extr_dir_identification_and_estim_3}. Similar qualitative behavior is observed under the mixture \HR~fit; therefore, we report only the results corresponding to the mixture logistic fit.
For the model in~\eqref{equa:simu}, both the correctly specified and misspecified model fits exhibit good performance, even for relatively small sample sizes, with satisfactory results from $n = 500$ onward.
In contrast, for the model perturbed as in~\eqref{equa:simu2}, performance shows a slight improvement as the sample size increases, however, remains inferior to that observed for Model~\eqref{equa:simu}. This behavior is expected. Indeed, for heavy-tailed random variables, the sum of two independent components is asymptotically equivalent to their maximum. Consequently, the random vector $\bX$ in~\eqref{equa:simu2} is asymptotically equivalent to a mixture model with extreme directions $\mathcal{J} = \mathcal{J}_A \cup \mathcal{S}_4$, where $\mathcal{J}_A$ are the extreme directions of the matrix $A$ in~\eqref{equa:matrix_A2} and $\mathcal{S}_4=\cbr{\cbr{1}, \cbr{2} , \cbr{3}, \cbr{4}}$.
Compared to Model~\eqref{equa:simu}, this formulation involves a larger number of parameters to estimate, which explains its comparatively weaker performance.
For the DAMEX algorithm, we follow the tuning recommendations in~\citet[Remark~7]{goix2017JMVA}, setting $\epsilon = 0.01$ and $k = n^{1/2}$. For the parameter $\mu$, we consider a grid of values $\cbr{0.01, \dots, 0.1}$ in steps of $0.01$ and select the value that minimizes the $\EDS$ score.
Overall, Algorithm~\ref{alg:extr_dir_identification_and_estim_3} yields better results than DAMEX. More specifically, for Model~\eqref{equa:simu}, Algorithm~\ref{alg:extr_dir_identification_and_estim_3} starts recovering the exact extreme directions associated to the matrix $A$ from a sample size of $n=1500$ in both the correct model and the misspecified model. In contrast, when using a sample size of the same order for DAMEX, the results were not satisfactory. Therefore, we consider a larger sample size of order $10^5$, as also adopted in the simulation study of~\citet[Section~5]{goix2017JMVA}.
We observe that the DAMEX consistently identifies the full set $\cbr{1,2,3,4}$ as an extreme direction, even though this is not supported by the matrix $A$ in~\eqref{equa:matrix_A2}. Moreover for Model~\eqref{equa:simu2}, the DAMEX does not identify the singletons as extreme directions.}

\amtwo{\paragraph{Computational complexity.} 
 In practice, changing the sample size $n$ or the tail fraction $k/n$ does not significantly affect the runtime of the algorithm.
 Typically, it runs in only a few seconds on a standard computer with $6$ cores.
The first computationally demanding part of Algorithm~3 is the $K$-fold cross-validation step used to select the optimal penalization parameter $\lambda$ from a grid of candidate values, denoted $\Lambda$. For each $\lambda \in \Lambda$, this requires:
\begin{enumerate}
    \item Splitting the data into $K$ folds.
    \item For each fold, fit the estimator on the training set and compute the cross-validation error on the held-out fold. This involves minimizing the loss function in Equation~(3.1) using a quasi-Newton method, which requires evaluating the loss multiple times to approximate the gradient and the Hessian.
\end{enumerate}
To reduce the computational burden, we rely on two strategies. First, we use parallel computing to distribute the values of $\Lambda$ across multiple cores. Second, we implemented the loss function in \texttt{C} in order to speed up its evaluation. The second computationally demanding part of Algorithm~3 is the dimension $d$.
In our experiments, using a supercomputer with 15 CPU cores and 15~GB of total RAM, with dimension $d=4$, sample size $n=1000$, tail fraction $k/n=0.08$, and a grid $\Lambda$ containing 20 values, the full cross-validation procedure takes approximately 3 minutes. 
}

\section{Applications}
\label{sec:application}
\subsection{River discharge data}
\label{sec:river_discharge}
We illustrate our method on river discharge data from $31$ stations along the Danube River. The dataset consists of daily river discharge measurements in the summer months between 1960 and 2010. To mitigate temporal dependence, the data were declustered in \citet{engelke2024package}, resulting in approximately seven to ten observations per year and yielding \( n = 428 \) observations in total. The declustered dataset is available in the \texttt{GraphicalExtremes} package in \texttt{R}; see~\citet{engelke2024package}.

To reduce the dimensionality and lower the computational cost of identifying extreme directions, we apply a spatial clustering technique. Specifically, we partition the stations into $K$ clusters using the Partitioning Around Medoids (PAM) algorithm~\citep{kaufman2009finding, bernard2013clustering}, adapted to threshold exceedances as in~\citet{kiriliouk2020climate}. As a pseudo-distance between two stations \( s \) and \( t \), we use
\begin{equation}
\label{equa:dist}
    \widehat{\dist}_{st} = \frac{1 - \hat{\chi}_{st}}{2 \cdot (3 - \hat{\chi}_{st})}, 
    \qquad 
    \hat{\chi}_{st} := 2 - \hat{\ell}_{n,k,st}(1,1),
\end{equation}
where $k/n$ represents the tail fraction and $\hat{\ell}_{n,k,st}$ the pairwise empirical stdf in~\eqref{eq:ellclassic} for the margin $\cbr{s,t}$.

\begin{figure}
    \centering
  \includegraphics[width=0.7\textwidth]{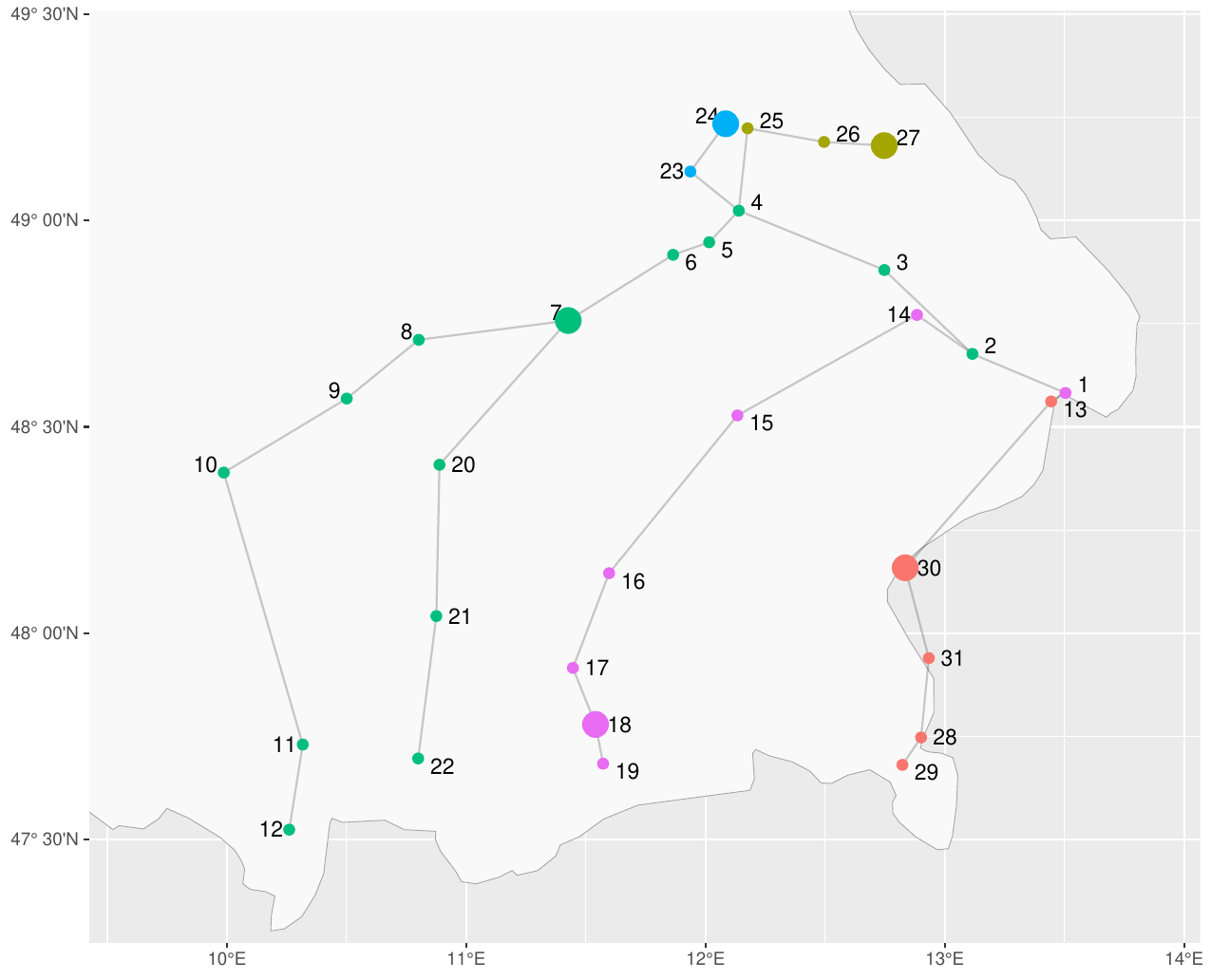} 
    \caption{\label{fig:danube_clustering} $31$ stations from the Danube  dataset clustered into $\mathcal K = 5$ groups using an adapted version of the PAM algorithm with the pseudo-distance $\widehat{\dist}$ in~\eqref{equa:dist} with tail fraction $k/n=0.1$. 
Stations with the same color belong to the same cluster, and within each cluster, the station represented by a larger-sized shape indicates the cluster center. }
\end{figure}

On the one hand, the number \( \mathcal K \) of clusters balances model simplicity (fewer clusters) and intra-cluster homogeneity. Various methods exist for selecting an appropriate value of \( \mathcal K \), such as minimizing within-cluster inertia~\citep{kaufman2009finding}. On the other hand, as explained in Section~\ref{sec:simu_study}, the choice of the tail fraction $k/n$ is a trade-off between bias and variance. Here, we simply choose $\mathcal K=5$ and $q = 0.05$ which gives spatially coherent results as displayed in Figure~\ref{fig:danube_clustering}. The cluster centers are given then by the five stations $\cbr{7 , 18 , 24 , 27 , 30}$. 

Consider the $d= \mathcal K =5$ stations of interest to be the cluster centers.
The data $\bX_i=(X_{i1} , \ldots , X_{id })$ representing the time series of river discharge over the five stations are assumed to be independent and identically distributed for $i=1,\ldots, n$ and we will identify their associated extreme directions using Algorithm~\ref{alg:extr_dir_identification_and_estim_3}.

\paragraph{Choice of tuning parameters.} Before estimating the extreme directions associated with the five central stations, we first address the choice of the tuning parameters: the tail fraction \( k/n \), the penalization exponent \( p \), the penalization coefficient \( \lambda \) and the grid points $\bo{c}_1,\ldots, \bo{c}_q \in [0,\infty)^5$. As noted in the simulation study, the choice of the penalization exponent \( p \) has limited impact provided that the penalization parameter \( \lambda \) is appropriately selected. We simply choose $p=0.4$, although, in terms of results, other values of the penalization exponent yield similar findings. Guided by the results in Section~\ref{sec:simu_study} and given the sample size, $n=428$, we set the tail fraction to \( k/n = 0.1 \). The penalization parameter \( \lambda \) is selected via $5$-fold cross-validation procedure as described in Algorithm~\ref{alg:cross_validation} over a grid of values. This grid is constructed in a greedy iterative way such that the optimal value $\lambda^{*}$ lies roughly in the middle of the range: we start with a grid \( \{ 0.1, 0.2 ,  \ldots, 10 \} \), ranging from $0.1$ to $10$ in steps of $0.1$, we then find the optimal value with respect to the $5$-fold cross-validation, and construct a new grid centered around this value. We repeat this process until results are stable. This yields a grid of values, \( \{ 0.01, 0.02 ,  \ldots, 1 \} \), ranging from $0.01$ to $1$ in steps of $0.01$. Note that given the limited sample size, $n=428$, we perform $5$-fold cross-validation, instead of $10$-fold cross-validation, as in Section~\ref{sec:simu_study}, where the sample size was larger. Finally, we select points that can be generated in a $5$-dimensional space using values $\cbr{1/4, \, 1/3 ,\,  1/2, \, 3/4 , \, 1}$  where each point satisfies the condition of having either two or three non-zero components. This yields $q= 1500$ evaluation points.

\paragraph{Results.} At each step of Algorithm~\ref{alg:extr_dir_identification_and_estim_3}, we fit a mixture logistic model to the five-dimensional river discharge dataset. While in the simulation study, the initial dependence parameter $\alpha$ of the mixture logistic model was randomly initialized within the interval $[0,1]$, here we set a fixed initial value $\alpha = 0.5$.  The results are summarized in Table~\ref{tab:extreme_directions_combined} (left). The algorithm identifies five extreme directions, which appear spatially consistent with the stations indicated on the map in Figure~\ref{fig:danube_clustering}. 

As indicated in Table~\ref{tab:extreme_directions_combined} (left), each extreme direction is associated with a weight, see \citep[Proposition~5.1]{mourahib2024multivariate}. Specifically, for a mixture logistic model with column coefficients $\rbr{\bo{a}_{\bigcdot J} : J \in \mathcal{J}}$, where $\mathcal{J}$ is the set of extreme directions (equivalently, the signature of the model) and a dependence parameter $\alpha \in [0,1]$, the scalar 
\begin{equation}
    \label{equa:weight}
    w_{\tilde{J}} = \frac{ \left(\sum_{j \in \tilde{J}} a_{j\tilde{J}}^{1/\alpha}\right)^{\alpha}}{\sum_{J \in \mathcal{J}}\left(\sum_{j \in J} a_{jJ}^{1/\alpha}\right)^{\alpha}}
\end{equation}
can be interpreted as the weight corresponding to the extreme direction $\tilde{J} \in \mathcal{J}$.

Results make sense given the hydrological nature of the dataset. The standard extreme direction $\cbr{7, 18, 24, 27, 30}$ can be explained by the fact that extreme river discharges are often caused by large-scale precipitation during the summer months; see~\citet{bohm2006flood}. The other non-standard extreme directions can be divided into two clusters.
The first cluster corresponds to the extreme directions $\cbr{24}$, $\cbr{24, 27}$, and $\cbr{7, 24}$, with stations 7, 24, and 27 located along Bavarian rivers. Stations 24 and 27 are geographically close and often experience the same weather systems. However, the presence of the extreme direction $\cbr{7, 24}$ is surprising, since the two stations receive water from different sources; see~\citet[Figure~9]{engelke2025extremes}. This might explain why the extreme direction $\cbr{7, 24}$ is detected with the lowest weight.
The second cluster corresponds to the extreme direction $\cbr{7, 18, 30}$, with stations 7, 18, and 30 located along Alpine tributaries (southern Germany and western Austria). This can be explained by the fact that these three stations receive water from the same region, flowing from south to north; see~\citet[Figure~9]{engelke2025extremes}.

\paragraph{\aktwo{Comparison to DAMEX and goodness of fit.}}  The DAMEX algorithm introduced in \citet[Algorithm~1]{goix2017JMVA} identifies only the extreme direction $\{7, 18, 24, 27, 30\}$, even when varying its tuning parameters. This extreme direction was also identified by our algorithm with the largest weight (see Table~\ref{tab:extreme_directions_combined} (left)). Our extreme direction identification method thus appears capable of detecting \amtwo{non-standard} extreme-direction structures. However, it is computationally more expensive than the DAMEX algorithm. 

To assess the fit of the mixture logistic model to the Danube dataset, we compare the empirical extremal correlation $\hat{\chi}_{st}$ in~\eqref{equa:dist} with the estimated extremal correlation for each pair of stations. For a mixture logistic model with estimated matrix $\hat{A} \in [0,1]^{d \times r}$ and dependence coefficient $\hat{\alpha}$, the extremal correlation between two margins $s$ and $t$ is
\aktwo{\begin{equation}
\label{equa:fit_chi}
    \hat{\chi}_{st , \, \est} = 2 - \sum_{k=1}^r \rbr{\hat{a}_{sk}^{1/\hat{\alpha} } + \hat{a}_{tk}^{1/\hat{\alpha}}}^{\hat{\alpha}}. 
\end{equation}}

Figure~\ref{fig:sidebyside_extcorr} (left) compares the empirical and fitted extremal correlations. Here we use $p=0.4$ and $k/n=0.1$, noting that other choices of tuning parameters yield qualitatively similar results. The agreement is strong, as most points lie close to the line $x=y$. Note that the evaluation process is entirely separate from the model fitting.

\subsection{Financial industry portfolios}
\label{sec:portfolios_app}

We consider the value-weighted returns for $d=10$ industry portfolios, downloaded from the Kenneth French Data Library\footnote{See~\url{https://mba.tuck.dartmouth.edu/pages/faculty/ken.french/Data_Library/det_10_ind_port.html} for more details about the data and the definition of each portfolio.}: “$1 = $ Nondurables”, “$2 = $ Durables”, “$3 = $ Manufacturing”, “$4 = $ Energy”, “$5 = $ HiTech”, “$6 = $ Telecom”, “$7 = $ Shops”, “$8 = $ Health”, “$9 = $ Utilities”, “$10 = $ Other”. The individual stocks that make up the five industry portfolios are taken from all listed firms on the {NYSE}, {AMEX}, and {NASDAQ}. Data is available between July 1926 and March 2025, leading to a sample of size $n=51\,922$. 

\paragraph{Data pre-processing.} We convert the standard returns to log-returns and multiply by $(-1)$ since we are interested in extreme \emph{losses}. Let $\bX_t = (X_{t1}, \ldots, X_{td})$ denote the negative log-returns of the five portfolios at time \(t = 1,\ldots,n\). In order to obtain a time series without temporal dependence, we fit a $\GARCH(1,1)$ model marginally. 
More precisely, for each portfolio $j=1,\ldots,10$,  
\begin{equation}
\label{equa:GARCH}
  X_{tj} = \mu_{tj} + \sigma_{tj} \varepsilon_{tj}, 
  \qquad t=1,\ldots , n,
\end{equation}
where \aktwo{$\sigma_{tj}^2$ is the conditional variance (volatility), and $\varepsilon_{tj} \overset{\textnormal{iid}}{\sim} \mathcal{N}(0,1)$}. For more details about the GARCH model and its application to financial data, see for example~\citet{francq2019garch}. We continue with the fitted residuals
\begin{equation*}
\hat{\varepsilon}_{tj} = \frac{X_{tj} - \hat{\mu}_{tj}}{\hat{\sigma}_{tj}}, \qquad t = 1,\ldots,n.    
\end{equation*}
We apply the Ljung--Box test for serial correlation to each of the $d=10$ marginal residual series $\hat{\varepsilon}_{1j}, \ldots,\hat{\varepsilon}_{nj}$, using lags $1$ through $10$. In every case, the test fails to reject the null hypothesis of no autocorrelation, suggesting that the residuals are serially uncorrelated. 

\paragraph{Choice of tuning parameters.} Similarly to Section~\ref{sec:river_discharge}, we set the penalization exponent to $p=0.4$. Other values of the penalization exponent yielded similar findings. The sample size ($n=51\,922$) of the industry portfolios dataset is much larger than the one of the Danube dataset ($n=428$) in Section~\ref{sec:river_discharge}. Therefore, in this section, we set the tail fraction $k/n$ to a smaller value $k/n = 0.05$. The penalization parameter \( \lambda \) is selected via the $10$-fold cross-validation procedure described in Algorithm~\ref{alg:cross_validation} over a grid of values. As explained in Section~\ref{sec:river_discharge}, this grid is chosen in a greedy iterative way until results stabilize. This yields a grid of values, \( \{ 0.001, 0.002 ,  \ldots, 0.02 \} \), ranging from $0.001$ to $0.02$ in steps of $0.001$. Finally, we select points that can be generated in a $10$-dimensional space using values $\cbr{0,  1/3 ,\,  1/2, \, 1}$  where each point has two non-zero components which yields $q= 405$ evaluation points.
 
\paragraph{Results.} As in Section~\ref{sec:river_discharge}, at each step of Algorithm~\ref{alg:extr_dir_identification_and_estim_3} we fit a mixture logistic model (Example~\ref{example:mix_log}) to the five industry portfolio and set the initial value for the dependence parameter to $\alpha = 0.5$. The results are summarized in Table~\ref{tab:extreme_directions_combined} (right). Unlike the DAMEX algorithm, our approach identifies multiple extreme directions, including non‑standard ones. Among these, the standard extreme direction $\{1,\ldots,10\}$ receives the highest weight, corresponding to the unique extreme direction detected by DAMEX, even when varying its tuning parameters.  Again, this emphasizes that our extreme direction identification method is capable of detecting sparser extreme-direction structures. The standard extreme direction $\{1, \ldots, 10\}$ corresponds to the one assigned the highest weight, which is intuitive given that the portfolios are composed of stocks from U.S.-based exchanges (NYSE, AMEX, and NASDAQ). As such, losses are not confined to a single sector but rather spread across the entire U.S. economy. Some of the other extreme directions can also be meaningfully interpreted. For instance, “$1 = $ NonDurables”, includes stocks from sectors such as Food and Tobacco, while “$10 = $ Other”, comprises sectors like Transportation and Entertainment; these two belong to the same extreme direction, suggesting similar tail-risk behavior. Likewise, “$6 = $ Telecom” (including Telephone and Television Transmission), and “$7 = $ Shops” (including Repair Shops), also fall within the same extreme direction. However, some extreme directions are more challenging to interpret without deeper economic insight. This includes singleton directions such as “$2 = $ Durables”, $4 = $ Energy”, “$6 = $ Telecom”, which may reflect more nuanced or isolated sector-specific behaviors. Finally, the standard extreme direction $\{1, \ldots, 10\}$ prevents the existence of a negative dependence structure between every pair of distinct sectors.

\begin{table}[t]
\centering
\renewcommand{\arraystretch}{1.2} 
\begin{minipage}[t]{0.48\textwidth}
\centering
\begin{tabular}{lc}
\toprule
Extreme directions & Weights \\
\midrule
$\cbr{7 , 18 , 24 , 27 , 30}$ & $0.49$ \\
$\cbr{7 , 18 , 30}$ & $0.17$ \\
$\cbr{24}$ & $0.15$ \\
$\cbr{24 , 27 }$ & $0.14$ \\
$\cbr{7 , 24}$ & $0.06$ \\
\bottomrule
\end{tabular}
\end{minipage}
\hfill
\begin{minipage}[t]{0.48\textwidth}
\centering
\begin{tabular}{lc}
\toprule
Extreme directions & Weights \\
\midrule
$\cbr{1,\ldots,10}$ & $0.52$ \\
$\cbr{7,8}$ & $0.10$ \\
$\cbr{6}$ & $0.09$ \\
$\cbr{1,9,10}$ & $0.09$ \\
$\cbr{4}$ & $0.08$ \\
$\cbr{1,2,3,6,7}$ & $0.07$ \\
$\cbr{2}$ & $0.02$ \\
\bottomrule
\end{tabular}
\end{minipage}
\vspace{1em}
\caption{Extreme directions identified by Algorithm~\ref{alg:extr_dir_identification_and_estim_3} for the Danube river discharge data at five stations, 
discussed in Section~\ref{sec:river_discharge} (left), and the ten industry portfolios: “$1 = $ Nondurables”, “$2 = $ Durables”, “$3 = $ Manufacturing”, “$4 = $ Energy”, “$5 = $ HiTech”, “$6 = $ Telecom”, “$7 = $ Shops”, “$8 = $ Health”, “$9 = $ Utilities”, “$10 = $ Other”, discussed in Section~\ref{sec:portfolios_app} (right), along with their associated weights as in~\eqref{equa:weight}. 
}
\label{tab:extreme_directions_combined}
\end{table}

\begin{figure}[ht]
    \centering
    \vspace{0.5cm}
        \includegraphics[width=0.4\linewidth]{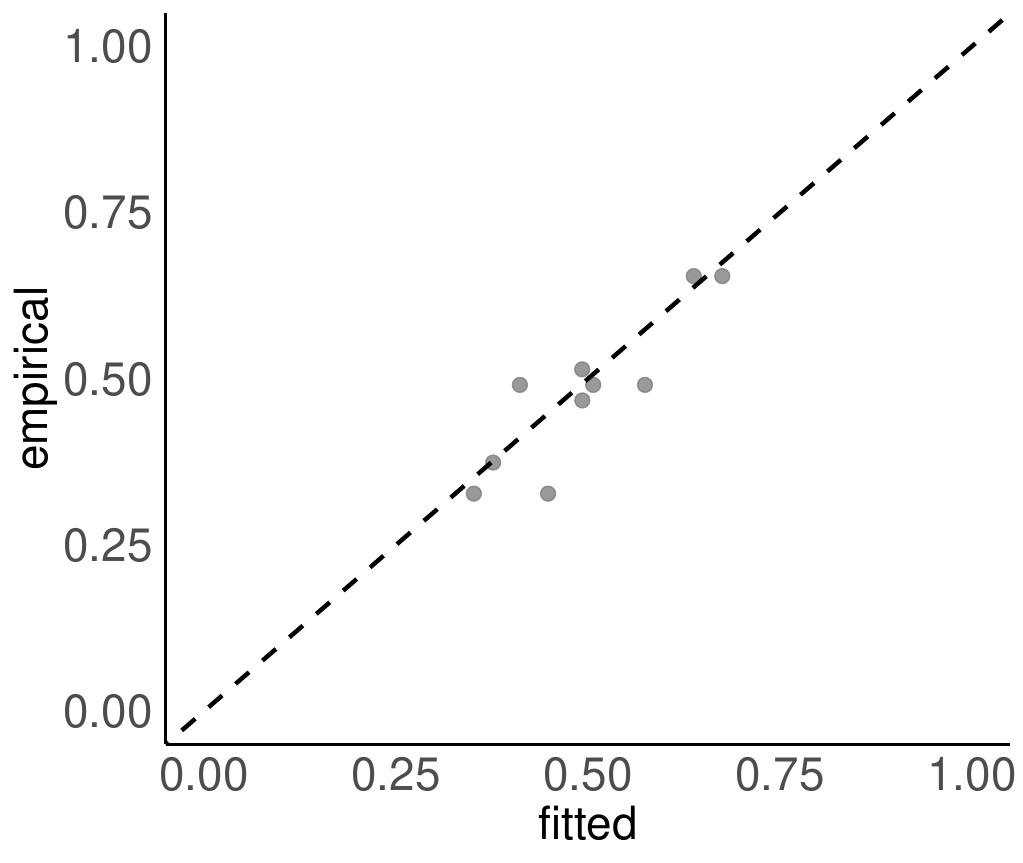}
        \hspace{1cm}
        \includegraphics[width=0.4\linewidth]{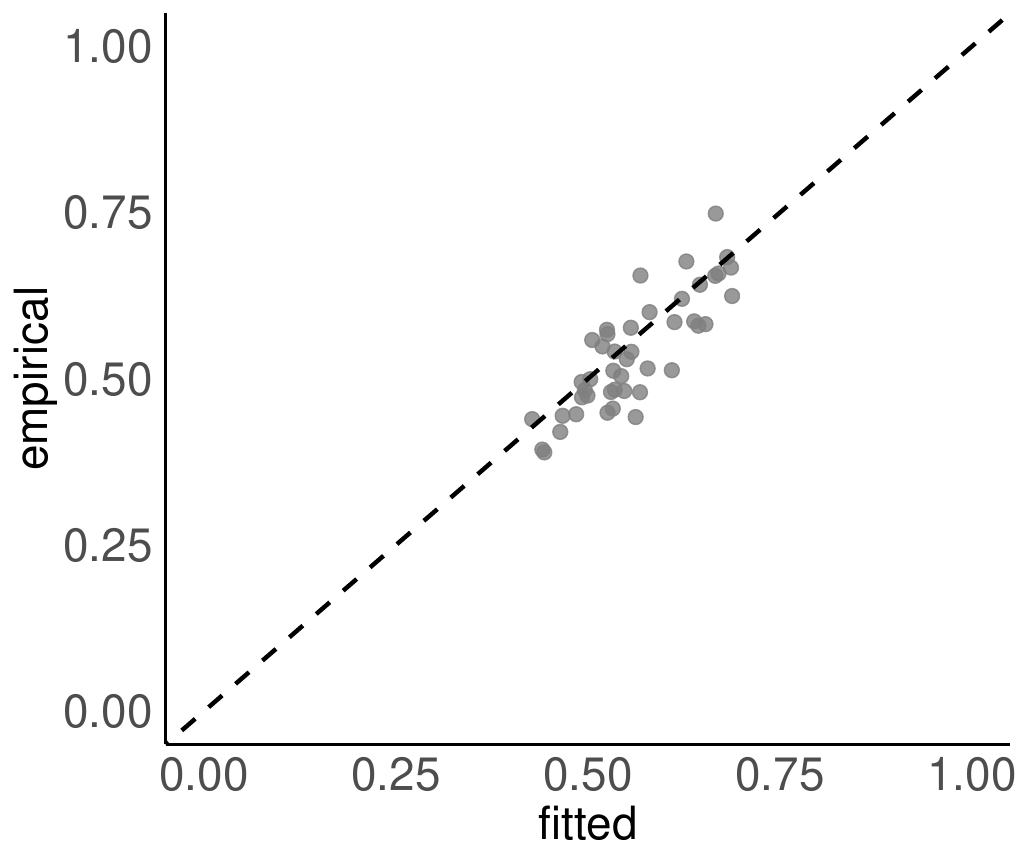}
    \caption{\amtwo{Empirical~\eqref{equa:dist} versus fitted~\eqref{equa:fit_chi} extremal correlations for each pair of margins. Left: results from the Danube dataset (Section~\ref{sec:river_discharge}) with $k/n = q = 0.1$. Right: industry portfolios (Section~\ref{sec:portfolios_app}) with $k/n = q = 0.05$. In both applications, the penalization exponent is fixed at $p = 0.4$.}}
    \label{fig:sidebyside_extcorr}
\end{figure}

To assess the goodness-of-fit of the mixture logistic model to the industry portfolios dataset, we compute for each pair $(s,t)$ both the empirical extremal correlation in~\eqref{equa:dist} and the fitted extremal correlation in~\eqref{equa:fit_chi}. Figure~\ref{fig:sidebyside_extcorr} (right) compares the empirical and fitted extremal correlations, computed both with $p=0.4$ and $k/n=0.05$. Other choices of tuning parameters yielded qualitatively similar results. The agreement is strong, as most points lie close to the line $x=y$.

\section{Conclusion}

Given a vector of $d$ risk factors, extreme directions refer to groups of factors $J \subseteq \cbr{1,\ldots,d}$ that are large simultaneously while those outside this group are not. \citet{mourahib2024multivariate} propose a construction method of multivariate generalized Pareto distributions, or equivalently max‐stable distributions, with multiple extreme directions. Extreme directions are a consequence of parameters of the mixture model being on the border of the parameter space \citep[Proposition~4.5]{mourahib2024multivariate}. Parameter estimation is therefore a challenging topic. We proposed a penalized least‐squares estimator~\eqref{eq:estimator} based on the estimator from~\cite{einmahl2018continuous}. 
\aktwo{We prove its consistency and propose} a data‐driven algorithm for extreme‐directions identification. Using a simulation study, we show that although the algorithm relies on a parametric assumption for the mixture model (e.g., mixture \HR~graphical model or mixture logistic model), the specific choice of the model is not crucial. In addition, the algorithm involves three tuning parameters; again through simulations, we demonstrate that the tuning essentially reduces to selecting two parameters: one that controls the proportion of data considered as extreme observations and the other that controls the penalization strength, chosen using $K$-fold cross-validation. We illustrated our algorithm on two real datasets, determining the sets of variables that form extreme directions.
We compare our approach with the DAMEX algorithm in~\citet{goix2017JMVA} and show that our algorithm appears capable of detecting non‐standard extreme directions.


\subsection*{Acknowledgments} The research of Anas Mourahib was supported financially by the \emph{Fonds de la Recherche Scientifique – FNRS}, Belgium (grant number T.0203.21). Computational resources have been provided by the supercomputing facilities of the Université catholique de Louvain (CISM/UCL) and the Consortium des Équipements de Calcul Intensif en Fédération Wallonie Bruxelles (CÉCI) funded by the Fond de la Recherche Scientifique de Belgique (F.R.S.-FNRS) under convention 2.5020.11 and by the Walloon Region.

The authors thank Sebastian Engelke for insightful discussions on the interpretation of the river discharge data analysis in Section~\ref{sec:river_discharge}, Michel Denuit for his input on the interpretation of the portfolio data in Section~\ref{sec:portfolios_app}, and Anne Sabourin for generously sharing her code and providing a clear explanation of the DAMEX algorithm.

\appendix 
\section{Mixture model simulation algorithm}
\label{app:algo}

\begin{algorithm}[H]
\caption{Simulation of a mixture model}
\label{alg:N_generate_Mix_log}
\hspace*{\algorithmicindent} \textbf{Input:}  integers $d,r \geq 1$,
matrix \( A = \rbr{a_{jk}}_{j \leq d, k \leq r} \in [0,1]^{d \times r} \) and max-stable random vectors $\bo{Z}_{\bigcdot 1} ,\ldots,\bo{Z}_{\bigcdot r} \in \reals^d$ as in Definition~\ref{def:mixture}. \\
\Comment{For simplicity, we consider the vector \( \bZ_{\bigcdot k} = (Z_{1k}, \ldots, Z_{dk}) \in \mathbb{R}^d \) for each \( k = 1, \ldots, r \). However, as discussed after Definition~\ref{def:mixture}, only the sub-vector \( \bZ_{J_k, k} \) is relevant.}\\
\hspace*{\algorithmicindent} \textbf{Output:}  mixture logistic model $\bM \in \reals^d$ in~\eqref{mixturemodel} with coefficient matrix $A$ and factors $\bo{Z}_{\bigcdot 1},\ldots,\bo{Z}_{\bigcdot r}$. 
\begin{algorithmic}[1]
\State Initialize a vector $\bM=(0,\ldots, 0) \in \reals^d$
\State Initialize a list \( \Tilde{Z}=\cbr{\Tilde{\bo{Z}}_1,\ldots, \Tilde{\bo{Z}}_r   } \), where each $\Tilde{\bo{Z}}_k \in \reals^d$ is initialized as a zero vector. 
\For{each \( k \in \{1, \dots, r\} \)}
    \State Define $J_k$ as in~\eqref{eq:Jk}, the $k$-th signature of the coefficient matrix $A$.
    \State Set $\Tilde{Z}_{k,j}$, the $j$-th component of $\Tilde{\bo{Z}}_k$, to \( a_{jk} Z_{jk}  \), for every $j \in J_k$.
\EndFor
\State Compute $M_j = \max \cbr{\Tilde{Z}_{j1} , \ldots , \Tilde{Z}_{jr}}$, for each $j \in \cbr{1,\ldots , d}$.
\State \Return \( \bM \)
\end{algorithmic}
\end{algorithm}

\section{Proof of Theorem~\ref{thm:consistency}}
\label{app:proof}
\begin{proof}[Proof of Theorem~\ref{thm:consistency}]
    By the consistency of the empirical stable tail dependence function \citep{drees1998best} as $k \to \infty$ and $k/n\to\infty$, we have $\hat{\ell}_{n,k}(\bc_m) \to \ell(\bc_m, \btheta_0)$ in probability for all $m \in \cbr{1,\ldots,q}$ as $n \to \infty$. In addition, $0 \le \ell(\bc_m; \btheta) \le \sum_{j=1}^d c_{mj}$ for all $m \in \cbr{1,\ldots,q}$ and $\btheta \in \Theta$, and $\mathcal{P}(A) \le d \cdot r$ for all $A \in [0, 1]^{d \times r}$. Since $\lambda \to 0$, it follows that
    \[
        \Delta_{n,k}
        := \sup_{\btheta \in \Theta} \left|
            \Loss_{n,k}(\btheta) - \Loss(\btheta)
        \right| \to 0
        \quad \text{in probability}
    \]
    as $n\to\infty$, where
    \begin{align*}
        \Loss_{n,k}(\btheta) &= 
        \sum_{m=1}^q \rbr{\hat{\ell}_{n,k} (\bc_m) - \ell(\bc_m ; \bo{\theta})}^2 + \lambda \mathcal{P}(A), \\
        \Loss(\btheta) &=
        \sum_{m=1}^q \rbr{\ell(\bc_m; \btheta_0) - \ell(\bc_m ; \bo{\theta})}^2.
    \end{align*}
    Clearly, $\Loss(\btheta_0) = 0$, while, by the injectivity assumption, $\Loss(\btheta) > 0$ as soon as $\btheta \ne \btheta_0$. Let $\eps > 0$ and write $K_\eps = \cbr{\btheta \in K : \| \btheta - \btheta_0 \| \ge \eps}$. By compactness of $K$ (and thus of $K_\eps$), we find
    \[
        \inf_{\btheta \in K_\eps}
        \Loss(\theta) =: \delta(\eps) > 0.
    \]
    On the event $\{\Delta_{n,k} < \delta(\eps)/2\}$, we have
    \[
        \forall \btheta \in K_\eps: \qquad
        \Loss_{n,k}(\btheta)
        > \Loss(\btheta) - \delta(\eps)/2
        \ge \delta(\eps)/2
        > 0 = \Loss(\btheta_0).
    \]
    It follows that, on the event $\{\Delta_{n,k} < \delta(\eps)/2\}$ we must have $\| \hat{\btheta}_{K} - \btheta_0 \| < \eps$. Since the probability of this event tends to zero and since $\eps > 0$ is arbitrary, the claim follows.
\end{proof}

\bibliographystyle{plainnat}
\bibliography{biblio}

\end{document}

%% file: newcommands.tex
\newcommand{\EDS}{\operatorname{ED-S}} 
\newcommand{\CV}{\operatorname{CV}}

\newcommand{\D}{\operatorname{D}}

\newcommand{\dist}{\operatorname{dist}}

\newcommand{\diffe}{\operatorname{diff}}
\newcommand{\argmin}{\operatornamewithlimits{\arg\min}}

\newcommand{\est}{\operatorname{est}}

\newcommand{\fs}{\operatorname{pr}}

\newcommand{\new}{\operatorname{new}}

\newcommand{\RRMSE}{\operatorname{SMSE}}
\newcommand{\GARCH}{\operatorname{GARCH}}
\DeclareMathOperator\symdif{\triangle}
\renewcommand{\ge}{\geqslant}
\renewcommand{\geq}{\geqslant}
\renewcommand{\le}{\leqslant}
\renewcommand{\leq}{\leqslant}
  
\makeatletter
\newcommand*\bigcdot{\mathpalette\bigcdot@{.5}}
\newcommand*\bigcdot@[2]{\mathbin{\vcenter{\hbox{\scalebox{#2}{$\m@th#1\bullet$}}}}}
\makeatother

\newcommand{\bU}{\boldsymbol{U}}

\newcommand{\bx}{\boldsymbol{x}}
\newcommand{\bX}{\boldsymbol{X}}
\newcommand{\bM}{\boldsymbol{M}}
\newcommand{\by}{\boldsymbol{y}}
\newcommand{\bY}{\boldsymbol{Y}}

\newcommand{\bZ}{\boldsymbol{Z}}

\newcommand{\bc}{\boldsymbol{c}}
\newcommand{\bo}[1]{\boldsymbol{#1}}
\newcommand{\binfty}{\boldsymbol{\infty}}
\newcommand{\bzero}{\boldsymbol{0}}
\newcommand{\bone}{\boldsymbol{1}}

\newcommand{\btheta}{\bo{\theta}}

\newcommand{\EE}{\mathbb{E}}

\newcommand{\reals}{\mathbb{R}}
\newcommand{\Rd}{\reals^d}

\newcommand{\EEJ}{\EE_{J}}

\newcommand{\hbt}{\hat{\bo{\theta}}}
\newcommand{\Loss}{\operatorname{Loss}}

\DeclarePairedDelimiterX{\norma}[1]{\lVert}{\rVert}{#1}

\newcommand{\bigar}{R}

\newcommand{\point}{\,\cdot\,}

\newcommand{\PP}{\operatorname{\mathsf{P}}}
\newcommand{\dto}{\rightsquigarrow}

\newcommand{\eps}{\epsilon}

\newcommand{\cbr}[1]{\left\{ {#1} \right\}} 
\newcommand{\rbr}[1]{\left( {#1} \right)}   
\newcommand{\sbr}[1]{\left[ {#1} \right]}   

\newcommand{\HR}{Hüsler--Reiss}
\newcommand{\FR}{Fr\'echet}
\newcommand{\extreme}{extreme direction}

\newcommand{\wrto}{w.r.t.\ }

\definecolor{navyblue}{rgb}{0.0, 0.0, 0.5}
\definecolor{darkcerulean}{rgb}{0.03, 0.27, 0.49}
\definecolor{burntorange}{rgb}{0.8, 0.33, 0.0}
\definecolor{britishracinggreen}{rgb}{0.0, 0.26, 0.15}
\definecolor{alizarin}{rgb}{0.82, 0.1, 0.26}
\definecolor{darkteal}{rgb}{0, 0.35, 0.35}

\definecolor{brightorange}{RGB}{255, 102, 0}

\definecolor{mediumpersianblue}{RGB}{0,103,165}

\newcommand{\amtwo}[1]{#1}  
\newcommand{\aktwo}[1]{#1}  